\documentclass{article}

\usepackage{arxiv}

\usepackage[utf8]{inputenc} 
\usepackage[T1]{fontenc}    
\usepackage{hyperref}       
\usepackage{url}            
\usepackage{booktabs}       
\usepackage{amsfonts}       
\usepackage{nicefrac}       
\usepackage{microtype}      
\usepackage{lipsum}		
\usepackage{graphicx}
\usepackage{natbib}
\usepackage{doi}
\usepackage{authblk} 
\usepackage{orcidlink} 
\usepackage{geometry}
\usepackage{graphicx}%
\usepackage{multirow}%
\usepackage{amsmath,amssymb,amsfonts}%
\usepackage{amsthm}%
\usepackage{tikz}
\usepackage{mathrsfs}%
\usepackage[title]{appendix}%
\usepackage{xcolor}%
\usepackage{textcomp}%
\usepackage{booktabs}%
\usepackage{algorithm}%
\usepackage{algorithmicx}%
\usepackage{algpseudocode}%
\usepackage{listings}%
\usepackage{float}
\usepackage{array}
\usepackage{tabularx}
\usepackage{bm}
\usepackage{adjustbox}
\title{A Physiologically-Constrained Neural Network Digital Twin Framework for Replicating Glucose Dynamics in Type 1 Diabetes}

\author{

\begin{tabular*}{\textwidth}{@{\extracolsep{\fill}}p{0.47\textwidth}p{0.47\textwidth}}
\parbox[c]{90mm}{\centering \textbf{Valentina Roquemen-Echeverri}\textsuperscript{1*} \orcidlink{0000-0002-5959-8022}} & \parbox[c]{90mm}{\centering \textbf{Taisa Kushner}\textsuperscript{2} \orcidlink{0000-0002-2726-0465}} \\
 \parbox[c]{90mm}{\centering\texttt{roquemev@ohsu.edu}} &  \parbox[c]{90mm}{\centering\texttt{taisa@galois.com}} \\
[1.5ex]
 \parbox[c]{90mm}{\centering\textbf{Peter G. Jacobs}\textsuperscript{1,3} \orcidlink{0000-0001-9897-4783}} &
\parbox[c]{90mm}{\centering\textbf{Clara Mosquera-Lopez}\textsuperscript{1} \orcidlink{0000-0003-1586-2490}} \\
\parbox[c]{90mm}{\centering\texttt{jacobsp@ohsu.edu}} &
\parbox[c]{90mm}{\centering\texttt{mosquera@ohsu.edu}} \\
\end{tabular*}
}

\affil{\textsuperscript{*}Corresponding author}
\affil{\textsuperscript{1} Artificial Intelligence for Medical Systems (AIMS) Lab, Department of Biomedical Engineering, Oregon Health \& Science University (OHSU), 3303 S Bond Ave, Portland, 97239, Oregon, United States of America}

\affil{\textsuperscript{2} Galois, Inc, 421 SW 6th Avenue \#300, Portland, 97204, Oregon, United States of America}

\affil{\textsuperscript{3}School of Chemical, Biological, and Environmental Engineering, Oregon State University, 1500 SW Jefferson Way, Corvallis, 97331, Oregon, United States of America}

\renewcommand{\shorttitle}{A Physiologically-Constrained Neural Network Digital Twin Framework}

\hypersetup{
pdftitle={A template for the arxiv style},
pdfsubject={q-bio.NC, q-bio.QM},
pdfauthor={David S.~Hippocampus, Elias D.~Striatum},
pdfkeywords={First keyword, Second keyword, More},
}

\begin{document}
\maketitle

\begin{abstract}

Simulating glucose dynamics in individuals with type 1 diabetes (T1D) is critical for developing personalized treatments and supporting data-driven clinical decisions. Existing models often miss key physiological aspects and are difficult to individualize. Here, we introduce physiologically-constrained neural network (NN) digital twins to simulate glucose dynamics in T1D. To ensure interpretability and physiological consistency, we first build a population-level NN state-space model aligned with a set of ordinary differential equations (ODEs) describing glucose regulation. This model is formally verified to conform to known T1D dynamics. Digital twins are then created by augmenting the population model with individual-specific models, which include personal data, such as glucose management and contextual information, capturing both inter- and intra-individual variability. We validate our approach using real-world data from the T1D Exercise Initiative study. Two weeks of data per participant were split into 5-hour sequences and simulated glucose profiles were compared to observed ones. Clinically relevant outcomes were used to assess similarity via paired equivalence t-tests with predefined clinical equivalence margins. Across 394 digital twins, glucose outcomes were equivalent between simulated and observed data: time in range (70-180 mg/dL) was 75.1$\pm$21.2\% (simulated) vs. 74.4$\pm$15.4\% (real; P<0.001); time below range (<70 mg/dL) 2.5$\pm$5.2\% vs. 3.0$\pm$3.3\% (P=0.022); and time above range (>180 mg/dL) 22.4$\pm$22.0\% vs. 22.6$\pm$15.9\% (P<0.001). Our framework can incorporate unmodeled factors like sleep and activity while preserving key dynamics. This approach enables personalized in silico testing of treatments, supports insulin optimization, and integrates physics-based and data-driven modeling. Code: \href{https://github.com/mosqueralopez/T1DSim_AI}{https://github.com/mosqueralopez/T1DSim\_AI}

\end{abstract}


\section{Introduction}\label{sec:introduction}
 
Type 1 Diabetes (T1D) is a chronic condition characterized by elevated glucose levels, resulting from the inability of the pancreas to produce enough insulin \cite{Katsarou_2017}. Thus, people with T1D must take exogenous insulin to enable their bodies to metabolize glucose. Achieving optimal glycemic control in T1D remains challenging as insulin therapy needs to be adjusted over time based on each individual's glucose response to meals, exercise, hormone cycles, stress, and other external disturbances \cite{Visentin_Dalla_Man_Kovatchev_Cobelli_2014, Hassan_Loar_Anderson_Heptulla_2006, Wellen_Hotamisligil_2005}. Advancements in diabetes technology including accurate, nonadjunctive continuous glucose monitoring (CGM) \cite{Castle_Jacobs_2016} and connected insulin pens and pumps, as well as automated insulin delivery (AID) and decision support systems (DSS) \cite{Tyler_Jacobs_2020,Tyler_Mosquera_2020} built on these devices, seek to alleviate this burden \cite{Breton_Farret_Bruttomesso_2012}. AID systems in particular have demonstrated significant improvements in glucose outcomes, increasing glucose time in range (TIR, 70-180 mg/dL) and reducing both time above range (TAR, \textgreater180 mg/dL) and time below range (TBR, \textless70 mg/dL) \cite{Wilson_2022}. Central to the development of AID and DSS have been virtual patient models, which are used as predictive models as well as the basis of \textit{in-silico} simulation environments used for pre-clinical validation of the systems' efficacy and safety. However, approximating the human glucoregulatory system remains difficult and most virtual patient models are lacking the individual-specific adjustments needed to ensure that \textit{in-silico} results translate to real-world individuals \cite{Mujahid_Contreras_Beneyto_Vehi_2024}.

Most existing T1D metabolic simulators are based on mechanistic models governed by ordinary differential equations (ODE) that model various factors such as carbohydrate intake and absorption, physical activity, insulin kinetics, and glucose-insulin dynamics \cite{Man_2014,Resalat_2019, Young_2023, Haidar_2013, Rashid_2019, Estremera_2022, Wilinska_2010, G_2023}. While these models provide an approximation of the human glucoregulatory system, they do not capture all physiological aspects impacting glucose dynamics and are difficult to fit to specific individuals. Identifying a personalized model that best matches a real-world person is oftentimes referred to identifying a digital twin for that person \cite{Tyler_Jacobs_2020}. There are a number of approaches for linking ODE models to individual data that have been described previously. Young et al. in \cite{Young_Dodier_2024} proposed an event-based approach to match a digital twin from a virtual population to real-world glucose traces before an exercise event based on insulin sensitivity and body weight \cite{ETD_young_2023}. The selected digital twin was used to simulate multiple interventions and select optimal recommendations specifically for enabling safe exercise in T1D. Cappon et al. \cite{G_2023} developed a two-stage digital twin-based simulation methodology that leverages observed glucose management data to identify a personalized model to approximate an individual's postprandial glucose response using Markov Chain Monte Carlo (MCMC). Then, the personalized model was used to simulate alternative insulin and carbohydrate therapies and evaluate their impact on glucose outcomes. MCMC is a robust approach to do model identification; however, it is time consuming and computationally expensive. Moreover, a more comprehensive evaluation of the proposed methodology on real-world data is lacking.

As an alternative to using ODE models to represent digital twins, a variety of groups have proposed data-driven methods for constructing digital twins. These data-driven approaches utilize artificial intelligence (AI) models, which can learn complex patterns from large datasets. However, these models are often not interpretable and risk mistaking patterns in data for causal relationships, unless designed with constraints \cite{Mujahid_Contreras_Beneyto_Vehi_2024, Kushner_Sankaranarayanan_Breton_2020, Erge_van_Oort_2022, Prendin_Pavan_Cappon_Del_2023}.

In recent years, there has been increasing interest in combining mechanistic or physics-based and data-driven AI-based modeling techniques. The domain knowledge built into mechanistic models is leveraged to inform the architectural design of AI models and to guide the learning process using aggregate supervision and constraints \cite{Rai_Sahu_2020, Wang_Yu_2023}. This hybrid modeling approach, also known as physics-informed machine learning, yields models that can accurately capture complex patterns from the data while adhering to physics principles. Hybrid modeling has been applied in various fields such as geophysics \cite{Yang_Wang_Sheng_Lin_Yang_2024}, epidemiology \cite{Wu_Gao_Xiong_Chinazzi_Vespignani_Ma_Yu_2021}, and fluid dynamics \cite{Wang_Kashinath_Mustafa_Albert_Yu_2020}. 

Hybrid approaches have been proposed for glucose prediction \cite{Zarkogianni_2014, Contreras_2017, Balakrishnan_2013,Zou_2024}, most of which predict future glucose over a fixed prediction horizon rather than being designed to simulate the dynamics of the glucoregulatory system states over time given simulation scenarios. These methods typically require historical data to contextualize predictions, and the hybrid component comes from using compartmental models to pre-process insulin kinetics and carbohydrate absorption before learning a function to predict glucose levels.

Herein we provide a novel hybrid modeling approach for constructing physiologically-constrained neural network (NN)-based digital twin models architected based on a mechanistic model, enabling conformance of the NN model with known physiologic constraints. We find these models are more accurate than comparator ODE-based mechanistic models in replicating real-world scenarios. See Appendix \ref{apx:related_work} for a detailed comparison of our approach with existing digital twin frameworks. Key contributions include:

\begin{enumerate}
    \item \textbf{Novel Model Architecture:} A population-level NN state-space model architected to be consistent with the ODEs describing the glucoregulatory system of individuals with T1D. This model is interpretable such that its consistency with physiology can be observed and formally verified (Section \ref{subsec:popModel}).
    \item \textbf{Formal Verification of Dynamics:} A methodology for verifying whether each sub-network of the population-level model replicates the dynamics present in the associated compartmentalized mechanistic model (Section \ref{subsec:conformance}).
    \item \textbf{Digital Twins:} A methodology for creating digital twins by combining the population-level model with individual-level models that augment predictions through integration of additional glucose management and contextual data to model inter- and intra-individual variability in glucose dynamics. By learning individual-level residuals directly from the rate of glucose change over time and dynamically adjusting glucose values at each simulation step, rather than learning the error for a given glucose prediction, our method ensures that the residuals remain independent of factors such as elapsed simulation time. We also created a large virtual population validated using real-world data from the T1D exercise initiative (T1DEXI) study (Section \ref{subsec:indModel}).
\end{enumerate}

We organize this paper as follows: Section \ref{sec:methods} describes our modeling and conformance verification approach, as well as the methodology to build physiologically-constrained NN digital twins. Section \ref{sec:results} reports simulation accuracy results. In Section \ref{sec:discussion}, we discuss our most relevant findings and the strengths and limitations of our work. Section \ref{sec:conclusion} concludes the paper. 

\section{Methods}\label{sec:methods}
\begin{figure}[h]
    \centering
    \includegraphics[scale=0.35]{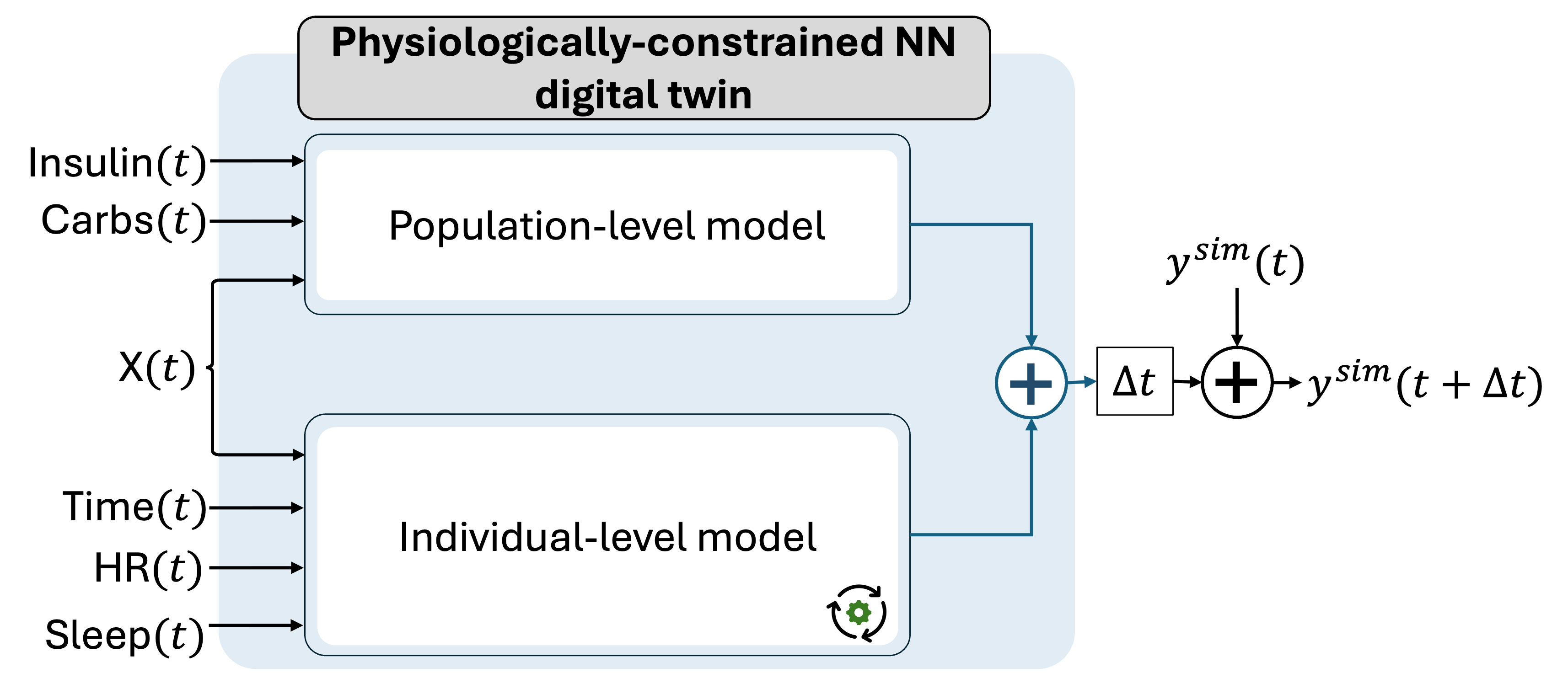}
    \caption{Overview of the physiologically-constrained NN digital twin framework for simulating glucose dynamics in type 1 diabetes. A digital twin consists of a population-level neural network state-space model and an adaptive individual-level neural network that models residual dynamics.
    X(t), HR(t), and $y^{sim}(t)$ represent the system's states, heart rate and glucose level, respectively, at time $t$}
    \label{fig:highLevelDiagram}
\end{figure}

The process for developing physiologically-constrained NN digital twins involves 3 steps: (1) developing a population-level NN state-space model of the glucose-insulin dynamics, (2) personalizing the population-level model for each individual; and (3) assessing their simulation accuracy. Our proposed framework for constructing digital twins is depicted in Fig. \ref{fig:highLevelDiagram}.

\subsection{Datasets}\label{subsec:datasets}

We used both simulated and real-world data to develop and test the physiologically-constrained NN digital twins. We used an open source metabolic simulator \cite{Resalat_2019} that enabled us to (1) observe all model states at any given time, (2) create a large and diverse dataset to capture the glucose dynamics under various real-world meal and insulin dosing scenarios, and (3) avoid potential errors found in free-living datasets, such as those related to self-reported events. Simulated data was used to develop the population-level model. Then, we used a real-world dataset for individual-specific augmentation of the population-level model (i.e., the creation of the digital twins).

To train the population-level model of the glucoregulatory system, we simulated glucose management data using a validated single-hormone simulator developed by Resalat et al. \cite{Resalat_2019} referred as $T1DSim_{ODE}$ (code available at \url{https://github.com/petejacobs/T1D_VPP}).

The real-world data came from the T1DEXI study, a large observational study that collected free-living glucose management data from 497 physically individuals with T1D across the United States (mean age $37\pm14$ years; HbA1c 6.6$\pm$0.8$\%$ [89$\pm$157 mg/dL]) using multiple daily injections (MDI), standard insulin pump, or AID therapies. The aim of the T1DEXI study was to investigate the effects of different types of exercise (i.e., aerobic, interval, and resistance) on glucose outcomes. During four weeks, study participants wore a CGM and a fitness tracker, and used an insulin delivery device and a smartphone-based app to log food intake, glucose management data and relevant life events. An Institutional Review Board approved the T1DEXI Study and electronic informed consent was obtained from each participant (dataset available for research purposes at \url{https://doi.org/10.25934/PR00008428}). The study had 2 phases: the initial pilot data collection \cite{Riddell_Li_2021} followed by the subsequent main data collection \cite{Riddell_Li_2023}.
We used the daily meal scenarios from the T1DEXI pilot phase to generate simulated data, and data from the T1DEXI main phase data to create individual-level models. As MDI data was self-reported, to ensure correct model identification and validation, we limit our analysis to participants using either open loop insulin pumps or AID systems. Specifics of this subset are provided in Appendix \ref{apx:dataset}.

\subsection{Population-level model}\label{subsec:popModel}
To capture complex temporal relationships, even when using shallow NNs, we designed a NN state-space model based on the physiology knowledge embedded into the $T1DSim_{ODE}$ simulator, an open-source simulator developed by our group that was successfully validated against a clinical dataset \cite{Resalat_2019}. This simulator is governed by a glucoregulatory model that consists of eight ODEs describing insulin kinetics (equations \ref{eq:eq1a} - \ref{eq:eq1c}), insulin dynamics (equations \ref{eq:eq1d} - \ref{eq:eq1f}), and glucose kinetics (equations \ref{eq:eq1g} and \ref{eq:eq1h}). In addition to capturing complex temporal relationships, the NN state-space model design allows us to verify physiological conformance by adhering to the foundational dynamics of the $T1DSim_{ODE}$. This alignment allows the model to maintain consistency with real-world physiological behavior while accommodating variability in compartments where the dynamics are less certain. Furthermore, this population-level model enables us to track the evolution of the system's states over time starting from any given initial state that encodes the history of the system, similar to conventional ODE-based models.

\begin{subequations}
\begin{align}
    \dot{S_1}(t)& = u_I(t) - \frac{S_1(t)}{t_{max}}  \label{eq:eq1a}\\
    \dot{S_2}(t)& = \frac{S_1(t)}{t_{max}} - \frac{S_2(t)}{t_{max}} \label{eq:eq1b}\\
    \dot{I}(t)& = \frac{S_2(t)}{t_{max}V_I} - k_eI(t) \label{eq:eq1c} \\
    \dot{X_1}(t)& = -k_{a1}X_1(t) + S_{f1}k_{a1}I(t) \label{eq:eq1d} \\
    \dot{X_2}(t)& = -k_{a2}X_2(t) + S_{f2}k_{a2}I(t) \label{eq:eq1e} \\
    \dot{X_3}(t)& = -k_{a3}X_3(t) + S_{f3}k_{a3}I(t) \label{eq:eq1f} \\
    \dot{Q_1}(t)& = -X_1(t)Q_1(t)-F^c_{01}-F_R+k_{12}Q_2(t) \label{eq:eq1g}\\
    & \quad +U_G(t)+EGP_0(1-X_3(t)) \nonumber \\
    \dot{Q_2}(t)& = X_1(t)Q_1(t) - k_{12}Q_2(t) - X_2(t)Q_2(t) \label{eq:eq1h}
\end{align}
\label{eq:compartment-ODEs}
\end{subequations}
 $S1$ and $S2$ represent the masses of insulin [mU/kg] in the subcutaneous compartment and an unobservable compartment, respectively. $u_I$ represents the rate of insulin infusion into the subcutaneous space [mU/kg/min]. $I$ represents the plasma insulin concentration [mU/L], and $t_{max}$, $V_I$ and $k_e$ are the time-to-maximum absorption [min], distribution volume [L/kg] and elimination rate [min$^{-1}$] of insulin, respectively.

$X_1$[min$^{-1}$], $X_2$ [min$^{-1}$] and $X_3$ [unitless] represent the effect of insulin on glucose distribution, disposal, and suppression of endogenous glucose production ($EGP$), respectively. $S_{f1}$[(mU$\cdot$L$\cdot$min)$^{-2}$], $S_{f2}$ [(mU$\cdot$L$\cdot$min)$^{-2}$] and $S_{f3}$ [(mU$\cdot$L$\cdot$min)$^{-1}$] are the insulin sensitivity factors, and are the parameters used for representing inter-individual insulin sensitivity variability of the glucoregulatory system in T1D.

$Q_1$ and $Q_2$ are the masses of glucose in the accessible (i.e., plasma) and non-accessible (i.e., peripheral tissue) compartments given in [mmol/kg], respectively. $EGP_0$ is the basal endogenous glucose production at a theoretical zero insulin concentration [(mmol/kg)/min]. $F_{01}^c$ and $F_R$ are the non-insulin mediated glucose uptake and the renal glucose clearance rate, respectively [mmol/kg/min]. $U_G$ represents the glucose absorption rate from meals [mmol/kg/min] (Equation \ref{eq:Ug}).
\begin{equation}
    U_G(t)=\frac{D_GA_G(t-t_0)e^{-\frac{t-t_0}{t_{max,G}}}}{t_{max,G}^2}
\label{eq:Ug}
\end{equation}
In Equation \ref{eq:Ug}, $t_{max,G}$ is the time-to-maximum appearance rate of glucose in $Q_1$ [min], $A_G$ is the carbohydrate bioavailability [unitless], $t_0$ is the meal onset time [min] and $D_G$ is the estimated carbohydrate intake [mmol/kg]. For \textit{in-silico} simulations, $D_G$ is converted from grams to mmol/kg to be compatible with the variables of the glucose kinetics model (Equation \ref{eq:eq1g}).

The population-level model called $T1DSim_{NN}^{P}$ is a 10-dimensional NN state-space model with the general form given in Equation \ref{eq:eq_AIpop}.
\begin{subequations}
\begin{align}
     \dot{\mathbf{x}}(t)& = \mathbb{\mathbf{N}}(\mathbf{x}(t), \mathbf{u}(t); \Theta^P) \label{eq:eq_AIpopa}\\
     y(t)& =Q_1(t) \label{eq:eq_AIpopb}
\end{align}
\label{eq:eq_AIpop}
\end{subequations}
In Equation \ref{eq:eq_AIpop}, $\mathbf{x}(t) = \{x^i(t)\} = 
\{S_1(t), S_2(t), I(t), X_1(t),\\X_2(t),X_3(t),Q_1(t),Q_2(t),C_1(t),C_2(t)\}$ is the set of system's states at time $t$; $\mathbf{u}(t)=$ \{$u_I(t)$,$u_{carb}(t)$\} is the set of system's inputs, $u_I(t)$ is the insulin infusion rate [U/h] and $u_{carb}(t)$ is the carbohydrate intake [g] at time $t$; and $\mathbb{\mathbf{N}}=\{\mathbb{N}_{fi}\}$ is a set of fully connected NNs parameterized by $\mathbf{\Theta^P}=\{\theta^P_i\}$, with i = 1,2,...,10. 

The architecture of the $T1DSim_{NN}^{P}$ is described by Equation \ref{eq:nn-compartments}. Note that the $T1DSim_{NN}^{P}$ includes the $C_1$ and $C_2$ compartments to explicitly model carbohydrate absorption following the model structure described by Hovorka et al. in \cite{Hovorka+Others/2004/Nonlinear}. This definition allows the data driven model states to represent the ODE model states in the glucoregulatory model including the compartment states, metabolic fluxes, and dependencies among the system's inputs and states. This approach increases the number of parameters compared to the $T1DSim_{ODE}$, which may lead to longer inference time. However, this trade-off provides the advantage of not only achieving a physiologically meaningful representation of the system but also establishing a framework for future extensions.
\begin{subequations}
\begin{align}
     \dot{S_1}(t)& = \mathbb{N}_{f1}(S_1(t),u_I(t)) \label{eq:eq2a}\\
     \dot{S_2}(t)& = \mathbb{N}_{f2}(S_1(t),S_2(t)) \label{eq:eq2b}\\
     \dot{I}(t)& = \mathbb{N}_{f3}(S_2(t),I(t)) \label{eq:eq2c}\\
     \dot{X_1}(t)& = \mathbb{N}_{f4}(X_1(t),I(t)) \label{eq:eq2d}\\
     \dot{X_2}(t)& = \mathbb{N}_{f5}(X_2(t),I(t)) \label{eq:eq2e}\\
     \dot{X_3}(t)& = \mathbb{N}_{f6}(X_3(t),I(t)) \label{eq:eq2f}\\
     \dot{Q_1}(t)& = \mathbb{N}_{f7}(X_1(t),X_3(t),Q_1(t),Q_2(t),C_2(t)) \label{eq:eq2g}\\
     \dot{Q_2}(t)& = \mathbb{N}_{f8}(X_1(t),X_2(t),Q_1(t),Q_2(t)) \label{eq:eq2h}\\
     \dot{C_1}(t)& = \mathbb{N}_{f9}(C_1(t),u_{carbs}(t)) \label{eq:eq2i}\\
     \dot{C_2}(t)& = \mathbb{N}_{f10}(C_1(t),C_2(t)) \label{eq:eq2j}
\end{align}
\label{eq:nn-compartments}
\end{subequations}
\subsubsection{Development dataset generation}
To develop a diverse dataset, we conducted 7-day simulations using a variety of real-world meals, insulin dosing, and initial glucose conditions. Simulations were started at midnight using the $T1DSim_{ODE}$ with population-level parameters as presented in \cite{Resalat_2019}, with details on scenarios described below: 
\begin{itemize}
    \item \textbf{Meal scenarios:} Daily meal scenarios were obtained from the pilot phase of the T1DEXI study \cite{Riddell_Li_2021}. Data was pre-processed to remove unrealistic scenarios caused by errors in self-reported data by selecting those that were within the $5^{th}$ and $95^{th}$ percentiles of the distribution of number of meals per day (1-9) and total carbohydrate intake per day (30-359 g). For any given 7-day simulation, 7 daily meal scenarios were selected. 
    \item \textbf{Insulin dosing:}  The insulin-to-carb ratio is simulated using the “1700 rule” of Davidson et al. \cite{Davidson_Hebblewhite_Steed_Bode_2008}. Multiple bolus timing scenarios were utilized including insulin given at meal onset, and delayed by 5 to 45 minutes (with 5-minute incremental steps). In addition to varying bolus timing, bolus size was varied to include under- and over-dosing. For all simulations, basal insulin was delivered every 5 minutes.
    \item \textbf{Initial glucose values:} Foreach 7-day meal scenario, five different initial glucose values were drawn from a normal distribution, with the mean and standard deviation calculated using CGM values at midnight from the participant corresponding to each meal scenario. The mean of the midnight CGM was 156$\pm$45 mg/dL.
\end{itemize}
The final \textit{in-silico} development dataset $\mathcal{D}^P = \{\mathbf{x}(t), \mathbf{u}(t)\}$, reflects 323,400 days of simulated glucose management data divided into 7-day simulation scenarios.
$\mathcal{D}^P$ was split into $\mathcal{D}^P_{train}$ (60\%), $\mathcal{D}^P_{validation}$ (20\%), $\mathcal{D}^P_{test}$ (20\%) subsets for model training, validation, and testing, respectively. The split was performed randomly, but with controls to ensure the same daily meal scenario appeared in no more than one subset, ensuring independence and avoiding data leakage. 

To improve data balance across rare events, we included additional traces with glucose less than 70 mg/dL, as well as those with delayed meal boluses to better represent glucose levels above 250 mg/dL. Specifically we supplement scenarios with TBR greater than 20\% or percentage time above 250 mg/dL greater than 40\%. Additionally, robust scaling was utilized to reduce the effect of outliers on the model training (see Appendix \ref{sec:robustScaler} for more details).

\subsubsection{Model training} \label{sssec:simulationTraining}

$T1DSim_{NN}^{P}$ was trained following the truncated simulation error minimization methodology described by Forgione and Piga in \cite{Forgione_2021}. Given $\mathcal{D}^P_{train}$, we created batches containing $B^P$ sequences of size $M$ with 75\% overlap between consecutive sequences. We used batch training meaning that model parameters were updated iteratively using the average loss calculated after processing a given batch to find the optimal set of parameters \textbf{$\Theta^P$} that minimized the loss function defined in Equation \ref{eq:eq_cost}.
\begin{equation}
    \mathcal{L}^P_{total} = \mathcal{L}^P_{fit} + \alpha \mathcal{L}^P_{consistency}
 \label{eq:eq_cost}
\end{equation}
In Equation \ref{eq:eq_cost}, $\mathcal{L}^P_{fit}$ is the loss associated with the measured glucose compartment $Q_1$, $\mathcal{L}^P_{consistency}$ is the loss associated with all system's states, and  $\alpha\ge0$ is a regularization parameter selected to balance $\mathcal{L}_{fit}$ and $\mathcal{L}_{consistency}$. The $\mathcal{L}^P_{fit}$ is defined as
\begin{equation}
    \mathcal{L}^P_{fit}=\frac{1}{B^PM}\sum_{b=0}^{B^P-1}\sum_{m=0}^{M-1}(y_{b,m}^{sim} - y_{b,m})^2\mathcal{P}(y_{b,m}^{sim},y_{b,m}),
\label{eq:eq_fit}
\end{equation}
where $B^P$ is the batch size, $M$ is the training sequence length, $y^{sim}$ is the simulated $Q_1$ sequence, and $y$ is the actual $Q_1$ sequence. $\mathcal{P}(y^{sim},y)$ is a penalty function defined in Equation \ref{eq:eq_pen} according to the penalty proposed by Del Favero et al. in \cite{Del_Favero_Facchinetti_Cobelli_2012}. $\mathcal{P}(y^{sim},y)$ was designed to penalize the errors made by the model on infrequent yet clinically significant events, such as hypoglycemia (glucose $<$70 mg/dL) and extreme hyperglycemia (glucose $>$250 mg/dL); both of which pose challenges in glucose management for individuals with T1D \cite{Cryer_2010, Kitabchi_Umpierrez_Miles_Fisher_2009}. The penalty values in $\mathcal{P}(y^{sim},y)$ were determined based on the ratio of the glucose risk function $\mathcal{R}_{BG}(g)=22.77(ln(g)^{1.084}-5.381)^2$ defined by Kovatchev et al. \cite{Kovatchev_2017} at glucose levels $g = $ 70 mg/dL ($\mathcal{R}_{BG}$(70) = 7.8) and $g = $ 250 mg/dL ($\mathcal{R}_{BG}$(250) = 22.4), which is about 3. This selection ensures that the errors are appropriately penalized based on the risk of over- or under-predicting glucose in critical ranges.
\begin{equation}
\mathcal{P}(y^{sim},y) = \left\{
\begin{array}{ll}
      2 & y\leq 70 \wedge y^{sim} > y \\
      6 & y\geq 250 \wedge y^{sim} < y\\
      1 & otherwise \\
\end{array} 
\right. 
\label{eq:eq_pen}
\end{equation}
The consistency loss $\mathcal{L}^P_{consistency}$ is as a weighted sum of the errors calculated for each of the $S$ compartments in the NN state-space model, and it is defined as
\begin{equation}
    \mathcal{L}^P_{consistency}=\sum_{i=0}^{S-1}\omega_i\mathcal{L}^{P,i}_{consistency},
\label{eq:eq_consistency}
\end{equation}
where $S=10$ is the number of compartments in the model, $w_i$ is the weight associated with compartment $i$, with $\sum_{i=0}^{S-1}w_i=1$. $\mathcal{L}^{P,i}_{consistency}$ is the loss associated with the $i^{th}$ compartment in the state-space model, and it is defined in Equation \ref{eq:eq_consistency_i}.
\begin{equation}
\begin{split}
    \mathcal{L}^{P,i}_{consistency}=\frac{1}{B^PM}\sum_{b=0}^{B^P-1}\sum_{m=0}^{M-1}(x^{i,sim}_{b,m}- x^i_{b,m})^2\\
      +\beta\sum_{b=0}^{B^P-1}\sum_{m=0}^{M-1}max[0,-(x^{i,sim}_{b,m}-min(x^i)]
    \label{eq:eq_consistency_i}
\end{split}
\end{equation}
In Equation \ref{eq:eq_consistency_i}, the first component of the consistency loss is the mean squared error (MSE) between $x^{sim,i}$,  the simulated $i^{th}$ state, and the actual state $x^i$. The second component is a soft constraint regularized by the parameter $\beta \ge 0$ added to prevent the states taking values that are less than those observed in the training set. Note that the $MSE^i = 0$ and $min(x^i) = 0$ for $i=9$ corresponding to the compartment $x^9 = C_1$ because there is no ground truth data for this compartment, which we introduced to explicitly model the delay in carbohydrate absorption. However, by making $min(x^i) = 0$, we are including a penalty to prevent $C_1$ to take negative values, which is not physiologically possible as it represents a negative meal amount. This is aligned with the interpretation of the $C_1$ compartment as the amount of carbohydrates on board, a quantity inherently non-negative.

\subsubsection{Network architecture}
For this work, we selected the simplest architecture for each sub-network $\mathbb{N}_{fi}$ which corresponds to a NN of one hidden layer with a rectified linear unit (ReLU) activation function. 
The batch size was $B^P$ = 128, sequence length $M$ = 5 hours with sampling period $\Delta m$ = $\Delta t$ = 5 minutes, regularization constants $\alpha=0.7$ and $\beta=0.08$, state error weights $w^i=0.08\overline{3}$ i=1,2,3,4,5,6,10, $w^7=0.208\overline{3}$, $w^8=0.1\overline{6}$, $w^9=0.041\overline{6}$, starting learning rate $\lambda^P = 10^{-3}$ scheduled to be reduced by $e^{-0.1}$ every epoch. 
The training process for the $T1DSim_{NN}^{P}$ model is detailed in the Algorithm \ref{alg:algo_training}. We used the training dataset $\mathcal{D}^P_{train}$ and gradient-based optimization using the Adam optimizer with recommended default parameters \cite{Kingma_Ba_2017}.

The optimal number of neurons per sub-network hidden layer, $\mathbb{N}_{fi}$, was found using Bayesian optimization \cite{bayesian_2014}, from an initial wide range of 4 to 256 neurons. The optimization target was to minimize, over 20 epochs, the balanced simulation error in Equation \ref{eq:eq_bayesOptError} on $\mathcal{D}^P_{validation}$, and this was performed with 50 exploration and 50 exploitation iterations. Our $\mathcal{D}^P_{validation}$ was split into sequences of size $M$ = 5 hours with no overlap. 
\begin{equation}
    \mathcal{L}^P_{BayesOpt}=\frac{1}{3}\sum_{j=1}^{3}\sqrt{\frac{1}{M}\sum_{m=0}^{M-1}(y^{sim}_{j,m}- y_{j,m})^2}
    \label{eq:eq_bayesOptError}
\end{equation}
In Equation \ref{eq:eq_bayesOptError}, $j = 1,2,3$ are the groups of sequences in $\mathcal{D}^P_{validation}$ with the following characteristics:
\begin{itemize}
    \item \textbf{Group 1:} Sequences with TIR greater than 70\% and TBR less than 20\%.
    \item \textbf{Group 2:} Sequences with TBR greater than 20\%.
    \item \textbf{Group 3:} The remaining sequences not included in Group 1 or Group 2.
\end{itemize}

\begin{algorithm}[h] 
\caption{$T1DSim_{AI}^{P}$ training process}
\label{alg:algo_training}
\begin{algorithmic}[1]
\Require{Training dataset $\mathcal{D}^P_{train}$ split into $n_{sequences}$ of size M, number of epochs $n_{epochs}$, batch size $B^P$, learning rate $\lambda^P$} 
\Ensure{Optimized $T1DSim_{AI}^{P}$ parameters $\Theta^P$}
\Statex
    \State \textbf{initialize} the $T1DSim_{AI}^{P}$ parameters $\Theta^P$
    \State \textbf{set} number of iterations
        \Statex\hspace{\algorithmicindent}$n_{iterations} \gets \frac{{{n_{epochs}}{n_{sequences}}}}{{B^P}}$
    \State\textbf{for} $j \gets 0$ to $n_{iterations}-1$                    
        \Statex\hspace{\algorithmicindent}\textbf{select} $B^P$ M-sized sequences to form a training batch
        \Statex\hspace{\algorithmicindent}\textbf{simulate} each training sequence in the batch
        \Statex \hspace{\algorithmicindent}\hspace{\algorithmicindent}\textit{Euler's ODE integration method}
        \Statex \hspace{\algorithmicindent}\hspace{\algorithmicindent}Initial states: $\mathbf{x}^0_{seq}$
        \Statex \hspace{\algorithmicindent}\hspace{\algorithmicindent}System's inputs: $\mathbf{u}_{seq}(t)$
        \Statex \hspace{\algorithmicindent}\hspace{\algorithmicindent}$\mathbf{x}^{sim}_{seq}(0) \gets \mathbf{x}^0_{seq}$
                \Statex\hspace{\algorithmicindent}\hspace{\algorithmicindent}\textbf{for} $m \gets 0$ to $M-1$                    
                \Statex \hspace{\algorithmicindent}\hspace{\algorithmicindent}\hspace{\algorithmicindent}$\dot{\mathbf{x}}^{sim}_{seq}(m) \gets \mathbf{\mathbb{N}}(\mathbf{x}^{sim}_{seq}(m),\mathbf{u}(m);\Theta^P)$
                \Statex \hspace{\algorithmicindent}\hspace{\algorithmicindent}\hspace{\algorithmicindent}$\mathbf{x}^{sim}_{seq}(m + \Delta m) \gets \mathbf{x}^{sim}_{seq}(m) + \dot{\mathbf{x}}^{sim}_{seq}(m) \Delta m$
                \Statex\hspace{\algorithmicindent}\hspace{\algorithmicindent}\textbf{end for}
        \Statex\hspace{\algorithmicindent} \textbf{compute}  $\mathcal{L}^P_{total}$ using Equation \ref{eq:eq_cost}
        \Statex\hspace{\algorithmicindent} \textbf{evaluate} the gradients $\nabla_{\Theta^P}\mathcal{L}_{total}$
        \Statex\hspace{\algorithmicindent} \textbf{update} $\Theta^P$
        \Statex\hspace{\algorithmicindent}\hspace{\algorithmicindent}$\Theta^P \gets \Theta^P-\lambda^P\nabla_{\Theta^P}\mathcal{L}^P_{total}$
        \Statex\textbf{end for}
\end{algorithmic}
\end{algorithm}

\subsection{Conformance Verification for Ensuring NN Models Match Known Physics}\label{subsec:conformance}
In order to verify that the dynamics of the learned models match known physiology, we leveraged our previously developed approach of verifying $\delta$-monotonicity through range estimation \cite{Kushner_Sankaranarayanan_Breton_2020} (Algorithm \ref{alg:algo_conformanceVerification}). This approach enables us to determine whether or not each sub-network $\mathbb{N}_{fi}$ defined in Equation \ref{eq:nn-compartments} replicate the same general dynamics (e.g., rates of change and direction of change) present in the associated ODE-based compartment model in Equation \ref{eq:compartment-ODEs}, while allowing variability on the individual-level model in terms of specific parameters. Specifically, we constructed input-output properties for each set of equations based on the partial derivatives of the corresponding ODE. If we take as an example Equation \ref{eq:eq1a}, $\frac{\partial{\dot{S_1}(t)}}{\partial{S_1}} = -\frac{1}{t_{max}}$; then, we have that the corresponding NN in Equation \ref{eq:eq2a} ($\mathbb{N}_{f1}$) must have the property that \emph{all else equal}, the output from Equation \ref{eq:eq2a} should monotonically decrease as $S_1$ increases. Similarly, the output of $\mathbb{N}_{f1}$ should increase as $u_I$ increases. This approach allows us to formally verify whether the NN models conform to known physiological properties, while providing flexibility in the specific rates of change observed.

\begin{algorithm} 
\caption{Neural network conformance verification}
\label{alg:algo_conformanceVerification}
\begin{algorithmic}[1]
\Require{Neural network $\mathbb{N}_{fi}$, minimum and maximum value of all $\mathbb{N}_{fi}$ inputs, input to be tested ($x_{test}$ or $u_{test}$), property to be tested ($\uparrow$: monotonic increase or $\downarrow$: monotonic decrease), $\delta$, minimum-viable change $\Delta x^i_{min}$}
\Ensure{Conformance verification result}
\Statex
    \State\textbf{create} two replicas of $\mathbb{N}_{fi}$ as $\mathbb{N}_{fi-left}$ and $\mathbb{N}_{fi-right}$
    \State\textbf{execute} Gurobi solver to find the objective values $objVal$ and $objVal_{\epsilon}$, which given to $\mathbb{N}_{fi}$ would result in the minimum value of $z_i$
    \State\textbf{evaluate} $\mathbb{N}_{fi}$ at $objVal$ and $objVal_{\epsilon}$
    \State\textbf{if} property to be tested = $\uparrow$
        \Statex\hspace{\algorithmicindent} $z_i \gets \mathbb{N}_{fi}(objVal) - \mathbb{N}_{fi}(objVal_{\epsilon)}$
    \Statex\textbf{else}
        \Statex\hspace{\algorithmicindent}$z_i \gets \mathbb{N}_{fi}(objVal_{\epsilon)} - \mathbb{N}_{fi}(objVal)$
    \Statex\textbf{end if}
    \State\textbf{if} $z_i < \Delta x^i_{min}$
        \Statex\hspace{\algorithmicindent}$\mathbb{N}_{fi}$ is conformant w.r.t $x_{test}$ or $u_{test}$
    \Statex\textbf{else}
        \Statex\hspace{\algorithmicindent}$\mathbb{N}_{fi}$ is non-conformant w.r.t $x_{test}$ or $u_{test}$
        \Statex\hspace{\algorithmicindent}$\Delta x^i_{critical} \gets z_i$
    \Statex\textbf{end if}
\end{algorithmic}
\end{algorithm}

\begin{figure}[H]
\begin{center}
\begin{tikzpicture}
\draw[fill=gray!30,draw=black,thick] (-3.5, 2) rectangle (-0.5,-1);
\draw[fill=gray!30,draw=black,thick] (3.5, 2) rectangle (0.5,-1);
\node at (-2,0.5) {\Large $\mathbb{N}_{f_{i=9-left}}$};
\node at (2,0.5) {\Large $\mathbb{N}_{f_{i=9-right}}$};

\node[rectangle,draw=gray](n0) at (-3.5,3)   { $x^1$};
\node[rectangle,draw=gray] (n1) at (-2.9,3)  { $x^2$};
\node at (-2.25,3)  { \Large $\cdots$};
\node[rectangle,draw=black,fill=red!20] (n2)  at (-1.45,3) { $x_{test}$};
\node[rectangle,draw=black,fill=red!20](n3)  at (-0.17,3)  { $x_{test}+\epsilon$};
\node at (0.9,3)  { \Large $\cdots$};
\node[rectangle,draw=gray] (n4) at (1.6,3)  { $x^{10}$};
\node[rectangle,draw=gray] (n5) at (2.25,3)  { $u_{I}$};
\node[rectangle,draw=gray,fill=gray!20] (n6) at (3.10,3)  { $u_{carbs}$};

\node[circle, fill=red!20] (nz)at  (0,-1.5) {\small $z_{i=9}$};

\path[->, line width=1.0pt] 
(n6.south) edge (3.45,2)
(n6.south) edge (-0.8,2)
(n2.south) edge [color=red] (-2.4,2)
(n3.south) edge [color=red] (1.95,2);

\path[->, line width = 1.5pt] (-2,-1) edge (nz)
(2,-1) edge (nz)
(nz) edge (0,-2.25);
\end{tikzpicture}
\end{center}
\caption{Scheme for checking $\delta$-monotonicity of a network $\mathbb{N}_{fi}$ with respect to a specific input location $x_{test}$ adapted from Kushner et al. \cite{Kushner_Sankaranarayanan_Breton_2020}. In this example, conformance verification is performed on $\mathbb{N}_{f9}$, $x_{test} = x^9 = C_1$, and $u_{carbs}$ is the other input to the network.}
\label{fig:delta-monotonicity-scheme}
\end{figure}
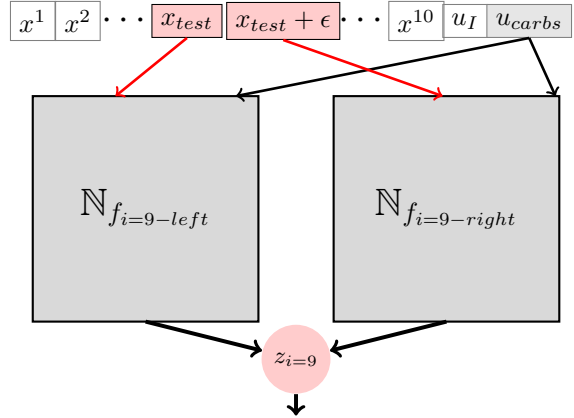

Following this rationale, we can define a set of properties for each NN in Equations \ref{eq:eq2a} - \ref{eq:eq2h}, based on their corresponding ODE in Equations \ref{eq:eq1a} - \ref{eq:eq1h}. For the compartments defined in equations \ref{eq:eq2i} and \ref{eq:eq2j}, the properties were defined based on the two-compartment ODE model described by Hovorka et al. in \cite{Hovorka+Others/2004/Nonlinear} for carbohydrate absorption. 

Using these properties as the ground truth, we set up two copies of the same network $\mathbb{N}_{fi-left}$ and $\mathbb{N}_{fi-right}$ in parallel, with all inputs but the one input being tested (e.g., $x_{test}$ or $u_{test}$) held equal, as shown in Fig. \ref{fig:delta-monotonicity-scheme}. If we were evaluating $x_{test}$, at the input location of $x_{test}$, $\mathbb{N}_{fi-left}$ gets $x_{test}$ and $\mathbb{N}_{fi-right}$ gets $x_{test} + \epsilon$, with $0 \le \epsilon  \le \delta$. We then construct a range estimation optimization problem to maximize (or minimize, when appropriate) the difference $z_i$ between the outputs of $\mathbb{N}_{fi-left}$ and $\mathbb{N}_{fi-right}$. This pushes the networks to extremes, allowing us to easily and formally identify any locations where the network dynamics break.  When testing for a monotonically decreasing property ($\downarrow$), we set $z_i = \mathbb{N}_{fi-right} - \mathbb{N}_{fi-left}$ and expect $z_i\leq0$. For a monotonically increasing property ($\uparrow$), we set $z_i = \mathbb{N}_{fi-left} - \mathbb{N}_{fi-right}$ and expect $z_i\leq0$.

The range estimation problem we just described asks for a conservative over-approximation of the output of a network given constraints on the inputs. For our approach, we cast the range estimation problem in a mixed integer linear programming (MILP) optimization framework, and solved it via the Gurobi solver~\cite{gurobi} to estimate the objective values $objVal$ and $objVal_{\epsilon}$ that would result in the minimum values of $z_i$. To validate the results of the MILP result, we checked the best-case solution (e.g., the example which falsifies the conformance problem) by evaluating the network $\mathbb{N}_{fi}$ with inputs $objVal$ and $objVal_{\epsilon}$  and calculating the actual value of $z_i$. In all cases, we desire $z_i\le0$. Thus, if we find a solution where $z_i>0$, we note the network $\mathbb{N}_{fi}$ is non-conformant. In essence, we took a falsification approach to solving the verification problem.

\subsubsection{Controlling for minimum-viable change in neural network compartments} 
When verifying conformance of the each $\mathbb{N}_{fi}$, the NNs can demonstrate dynamics which are below the discrepancy threshold for the Gurobi optimizer. To address this issue, we included a minimum-viable change $\Delta x^i_{min}$ based on the absolute value of smallest non-zero rate of change measured at the $i^{th}$ compartment over 1 time unit (i.e., $\Delta t = \Delta m = $ 5 min) in $\mathcal{D}^P_{train}$. We considered $\Delta x^i_{min}$ to be effectively zero based on physiology knowledge from the ODE-based glucoregulatory model and used it to determine whether or not a sub-network was conformant. Therefore, if we found a solution where $z_i< \Delta x^i_{min}$, we noted the network $\mathbb{N}_{fi}$ is conformant; otherwise, we reported the critical error $\Delta x^i_{critical}$ below which the network output would be considered partially conformant.
\subsection{Individual-level model}\label{subsec:indModel}
To construct the individual-level augmentation models, we developed a NN model $\mathbb{N}_{Ind}$, parameterized by $\theta ^I_k$, $k=1,..., K$, called $T1DSim_{NN,k}^{I}$ (Equation \ref{eq:eq_indLevelNNComp}). $K$ is the total number of digital twins to be constructed.
\begin{equation}
\begin{aligned}
    \dot{Q_1^I}(t) ={\mathbb{N}_{Ind}}({X_1}(t),{X_3}(t),{Q_1}(t),\\
    {Q_2}(t),{C_2}(t), \mathbf{u_{Ind}}(t);{\theta ^I_k}) 
\label{eq:eq_indLevelNNComp}
\end{aligned}
\end{equation}
We construct this model to be trained over time on the data generated by each individual using gradient descent, allowing the network to act as an additional compartment that captures the  inter- and intra-individual variability in glucose dynamics not modeled by the population-level model $T1DSim_{NN}^{P}$. $\mathbb{N}_{Ind}$ has as inputs the same states as the compartment ${Q_1}$ (Equation \ref{eq:eq2g}) along with the input vector $\mathbf{u_{Ind}}(t)$ which includes sleep efficiency, heart rate, and contextual time information (e.g., hour of day, and weekday vs. weekend) specific to a given individual. To capture complex individual patterns, we designed $\mathbb{N}_{Ind}$ as a multi-layer NN that consists of 3 fully connected hidden dense layers with 128, 64, and 32 neurons, respectively. Each hidden layer has a ReLU activation function. $T1DSim_{NN,k}^{I}$ is integrated in the simulation as shown in Equation \ref{eq:eq_indLevelNN}.
\begin{equation}
    y_{Ind}^{sim}(t + \Delta t) = Q_1(t) + (\dot{Q_1}(t)+\dot{Q_1^I}(t))\Delta t
    \label{eq:eq_indLevelNN}
\end{equation}
\subsubsection{Development dataset preprocessing}

The development dataset $\mathcal{D}^I_k = \{\mathbf{x}(t), \mathbf{u}(t),\mathbf{u_{Ind}}(t)\}$, $k=1,..., K$, was generated using data from the main phase of the T1DEXI study (see Appendix \ref{apx:dataset}). 

All the system's states ($\mathbf{x}(t)$) were simulated using the $T1DSim_{NN}^{P}$ during the training of the individual model, except for $x^7(t)=Q_1(t)$ which is the actual glucose measured by the CGM sensor. The system's input $u_I(t)$ was obtained from the insulin pumps and the carbohydrate intake $u_{carb}(t)$ from self-reported meal events. The self-reported meal events were confirmed by a validated meal detection algorithm \cite{MosqueraLopez_2023}. We first reviewed the predicted meals and included those that (1) had a bolus reported within a 90-minute window (for these, we used the estimated carbohydrate amount and the bolus timestamp) or (2) had a meal reported within a 90-minute window but no bolus reported (for these, we used the carbohydrate amount reported in that range with timestamp detected by the missed-meal algorithm). Next, we processed the remaining bolus events, checking for meal events within a 90-minute window and including them where applicable. Throughout this process, we ensured that no meal event was included more than once. Meals that did not meet any of these preprocessing criteria were discarded as unreliable. Initially, a total of 44,133 meals were reported across all datasets. After applying the described process, the number of meal events increased to 51,497. Of the initial events, only 11,205 retained their original size and timing without modification. The meal-detection algorithm was utilized to mitigate the problem whereby some individuals did not accurately report meal events, thereby improving the quality of the dataset. 

The individual input vector ($\mathbf{u_{Ind}}(t)$) included the following inputs:

\begin{itemize}
    \item \textbf{Timed-based contextual features:} Hour of day coded as $\cos \left( {2\pi \frac{{hour}}{{24}}} \right)$, $\sin \left( {2\pi \frac{{hour}}{{24}}} \right)$; and a binary variable to indicate whether or not the simulation instance corresponds to a weekend day.
    \item \textbf{Heart rate:} Change in heart rate measurements expressed in beats per minute (BPM), relative to the individual's baseline. The individual's baseline was estimated as the average heart rate during periods of rest, when no exercise was being performed.
    \item \textbf{Sleep efficiency:} Fraction of time within a 5-minute window spent sleeping.
\end{itemize}

The heart rate was scaled using a robust scaler as done for the states in the population-level model. The remaining inputs were constructed to be between 0 and 1 or between -1 and 1 (e.g., the hour feature), thus avoiding the need for scaling. 

$\mathcal{D}^I_k$ was divided into a training subset ($\mathcal{D}^I_{train,k}$) comprising the first two weeks of the study and a testing set ($\mathcal{D}^I_{test,k}$) corresponding to the remaining days. We chose a two-week training window because previous research has shown it to be the optimal duration for identifying individual-level glucose dynamics using machine learning, while also minimizing the risk of overfitting \cite{Herrero_Reddy_Georgiou_Oliver_2022}. If for a given participant, there were no insulin or carbohydrates reported in either of the two subsets, the participant was excluded from the virtual population.
\subsubsection{Model training}

The $T1DSim_{NN,k}^{I}$ models constructed for all digital twins were trained following the methodology outlined in section \ref{sssec:simulationTraining}. No modifications were made to the methods apart from those indicated in this section.

Given $\mathcal{D}^{I}_{train,k}$, we created batches containing 5-hour sequences, with each consecutive sequence overlapping 90\% with the previous one. The overlap ensures that the training data contains a larger number of samples, as a new sequence is generated by shifting the starting point by 10\% of the sequence length (e.g., 30 minutes for a 5-hour sequence). This approach increases the diversity of training samples by exposing the model to slightly shifted but highly similar input sequences, improving its robustness to variations in temporal alignment. The augmentation process is useful, as small shifts help capture subtle temporal patterns. Since we penalized each timestamp in the sequence, we required sequences to have 100\% of the CGM values present to be included in the training phase. 

The loss function for the training individual-level models is defined in Equation \ref{eq:eq_costInd}.

\begin{equation}
    \mathcal{L}^I_{total} = \mathcal{L}^I_{fit} +  \alpha^I\mathcal{L}^I_{consistency}
 \label{eq:eq_costInd}
\end{equation}

$\mathcal{L}^I_{fit}$ was defined as in Equation \ref{eq:eq_fit}. However, we redefined the penalty function as presented in Equation \ref{eq:eq_penInd} to penalize more heavily errors in the low glucose range. This penalty in the training process is important for ensuring that the digital twins are accurately representing glucose data in all ranges, including low glucose values which are important to model for any interventional system. High times below range are associated with high glucose variability, which is an increasingly relevant clinical marker of daily glucose control \cite{Wilmot_2019}. We searched for the penalty value for the condition $y\leq 70 \wedge y^{sim} > y$ in $\mathcal{P}^I(y^{sim},y)$ that would better approximate global TBR statistics for all digital twins in $\mathcal{D}^{I}_{train,k}$. 

\begin{equation}
\mathcal{P}^I(y^{sim},y) = \left\{
\begin{array}{ll}
      8.5 & y\leq 70 \wedge y^{sim} > y \\
      6 & y\geq 250 \wedge y^{sim} < y\\
      1 & otherwise \\
\end{array} 
\right. 
\label{eq:eq_penInd}
\end{equation}

$\mathcal{L}^I_{consistency}$ was defined as the mean squared error between the first-order differences of the simulated 5-hour sequence and the first order differences observed in the actual glucose data. This error term was included to avoid unrealistic changes within a 5-minute window in the simulation. And $\alpha^I\geq0$ is a regularization parameter selected to balance $\mathcal{L}^I_{fit}$ and $\mathcal{L}^I_{consistency}$ .

For training the individual-level models, we used a learning rate of $\lambda_{I} = 10^{-4}$ and a regularization constant $\alpha^I =10$. Each individual model was trained for 150 epochs.

\subsection{Simulation of real-world scenarios using physiologically-constrained NN digital twins}

\subsubsection{Estimating initial states} To initialize the simulator when only the initial CGM data was available, we developed an approach based on steady-state assumptions derived from the CGM readings at the start of the simulation ($t=0$). Our technique involves obtaining the set of initial states $\mathbf{x}(0)$ calculated when assuming $\dot{\mathbf{x}}(0) = 0$, given the model $T1DSim_{NN}^{P}$. This was achieved using a gradient-based optimization approach with the stochastic gradient descent optimizer. The number of epochs and the learning rate were empirically set to 10,000 and 0.1, respectively.

\subsubsection{Conducting simulations} Given the initial states $\mathbf{x}(0)$, and inputs $\mathbf{u}(t)$ and $\mathbf{u_{Ind}}(t)$, $t = 0, ... ,T$. A $T-$length simulation of the glucose dynamics of the $k^{th}$ digital twin parameterized by $\theta^I_k$ can be performed using the Euler's ODE integration method (Equation \ref{eq:eq_indLevelNN}).

\subsection{Evaluation of simulation accuracy}

\subsubsection{ODE-based digital twins used as control models} 

We compared our approach with other ODE-based methodologies. Similar published  digital twin frameworks in T1D are all based on mechanistic models governed by ODEs describing glucose-insulin dynamics \cite{Cappon_Facchinetti_2024}. Thus, we prioritized comparisons between our proposed methodology and the clinically validated ODE-based glucoregulatory model developed by Resalat et al. \cite{Resalat_2019}, as follows:
\begin{itemize}
    \item \textbf{ODE-based population-level model:} We simulated real-world scenarios using the population model parameters presented in \cite{Resalat_2019}.
    \item \textbf{Closest ODE-based digital twin from existing virtual population:} For each participant in the T1DEXI study included in our analysis, we identified the best digital twin from the population of 99 virtual patients described in \cite{Resalat_2019} by selecting the twin that over the first two weeks of the T1DEXI study minimized the error used in the loss function optimized during the individual-model optimization phase (Equation \ref{eq:eq_costInd}).
    \item \textbf{Bayesian optimization-based digital twin:} Similar to the ReplayBG twinning technique by Cappon et al. \cite{G_2023}, we used Bayesian optimization to find the optimal subset of the ODE-based model parameters related to insulin sensitivity (i.e.,$S_{f1}$, $S_{f2}$, and $S_{f3}$) that minimized simulation error for each participant in the T1DEXI study included in our analysis. Insulin sensitivity parameters are the most important parameters that capture inter-individual variability. We used Bayesian optimization \cite{bayesian_2014} instead of MCMC because this methodology requires substantial time and computational resources. MCMC is a sampling algorithm that construct Markov Chains to approximate distributions and is used for Bayesian inference to estimate posterior distribution of a model’s parameters. Bayesian optimization uses probabilistic modeling to optimize expensive functions and it is use for parameter tuning or black-box function optimization. In our work, the optimization target was to minimize the simulation error using the first two weeks of the T1DEXI study. 
    
\end{itemize}


\subsubsection{Statistical analysis of similarity between simulated and real-world glucose outcomes} 

The accuracy of the simulated glucose profiles was evaluated by comparing average outcome metrics between each digital twin and real-world data. Specifically, we assessed TIR, TAR, and TBR. Additionally, we computed clinically relevant risk indices, including the high and low blood glucose indices (HBGI and LBGI), as well as mean glucose (MG). These metrics are widely used to assess clinical performance and glucose control \cite{Battelino_2019}.

To ensure consistency with the model training data, all metrics were calculated over 5-hour sequences. This sequence length was chosen to reflect a realistic simulation window, capturing key physiological dynamics such as those associated with meals or physical activity.

We used paired equivalence t-tests to determine whether simulated and real-world outcomes were statistically equivalent within predefined clinically meaningful margins. The equivalence margins for each metric were: $\delta_{TIR} = 5\%$, $\delta_{TAR} = 5\%$, $\delta_{TBR} = 1\%$, $\delta_{HBGI} = 1$, $\delta_{LBGI} = 0.5$ and  $\delta_{MG} = 10$ mg/dL \cite{Battelino_2019,Battelino_2023, VillaTamayo_Colmegna_Breton_2024a}.
All tests were two one-sided paired t-tests (TOST) with a significance level of 0.050. Results are reported as mean $\pm$ standard deviation (SD), along with the larger of the two one-sided p-values.

\subsection{Additional experiments}

To further evaluate the utility and robustness of the digital twins, we conducted three complementary experiments:

\subsubsection{Contribution of individual-level inputs}

To assess the contribution of each component of the individual input vector $\mathbf{u_{Ind}}(t)$, we trained multiple versions of the digital twin model using different subsets of the input features. Specifically, we trained models using only one input at a time - heart rate, sleep efficiency, weekend day or time-based features—as well as a model without addition inputs (i.e., $\mathbf{u_{Ind}}(t)=\{\}$). They where compared with the NN-based digital twin proposed. 

Each model configuration was evaluated across three key physiological events known to impact glucose levels (i.e., meals, physical activity, and sleep). We used the same outcome metrics described above (TIR, TAR, TBR, HBGI, LBGI, and MG). This experiment aimed to determine whether incorporating all individual-level inputs provides additive value and improves simulation accuracy compared to models trained with partial or no individual-level information.

\subsubsection{Evaluation of digital twins behavior under variable insulin boluses and carbohydrate inputs}\label{sec:twinning effect}

We demonstrate the potential of the physiologically-constrained NN digital twins to replay various carbohydrate ratios by simulating three different meal scenarios (i.e., meal carbohydrate consumption of 30 g, 60 g, and 90 g) across all digital twins. For each meal scenario, we simulated insulin boluses for three different insulin-to-carbohydrate ratios (i.e., 10 g/U, 20 g/U, and 30 g/U). These simulations qualitatively illustrate that the digital twins can, for example, be used to adjust the optimal insulin-to-carbohydrate ratio for each individual based on improving the postprandial glucose response.

\subsubsection{Evaluation as a predictive model}

Although digital twins are designed for scenario-based simulation (i.e., forecasting glucose trajectories under varied conditions), we also evaluated their performance in a traditional predictive setting (i.e., forecasting glucose values over a short-term prediction horizon). Specifically, we compared the performance of the digital twins against state-of-the-art glucose prediction models \cite{Kushner_Sankaranarayanan_Breton_2020, Prendin_Pavan_Cappon_Del_2023, Kushner_Breton_Sankaranarayanan_2020} using a 30-, 60- and 120-minute prediction horizon. 

To perform the prediction, we used the glucose value at time 0 of a given scenario as the initial condition and simulated the glucose trajectory over the following hour using the same scenario. The predicted glucose value at time 30, 60 and 120 minutes was then compared to the actual value. For comparison, we used point-wise root mean squared error (RMSE) as the evaluation metric. This comparison helps establish the digital twins’ capabilities not only as simulators but also as effective short-term glucose predictors.

\section{Results}\label{sec:results}

Table \ref{table:best_archs} shows the number of hidden units for each sub-network that minimized validation error for the population-level model $T1DSim_{NN}^{P}$ and the individual-level models $T1DSim_{NN,k}^{I}$.

\begin{table}[ht]
\centering
    \caption{Number of units in the hidden layers of subnetworks in the population- and individual-level models}
\begin{tabular}{p{6mm}p{6mm} p{6mm}p{6mm} p{6mm}p{6mm} p{9mm} p{9mm}}
\toprule
\multicolumn{8}{c}{\textbf{Population-level Model ($T1DSim_{NN}^{P}$)}}  \\ \addlinespace[0.2em]  
\midrule 
\multicolumn{2}{c}{\parbox[c]{12mm}{\centering\textbf{Glucose\\kinetics}}} & 
\multicolumn{2}{c}{\parbox[c]{12mm}{\centering\textbf{Insulin\\kinetics}}} & 
\multicolumn{2}{c}{\parbox[c]{12mm}{\centering\textbf{Insulin\\dynamics}}} & 
\multicolumn{2}{c} {\parbox[c]{18mm}{\centering\textbf{Carbohydrate\\absorption}}} \\ \addlinespace[0.4em]  
$\dot{Q_1}$  & 248   & $\dot{S_1}$    & 174 & $\dot{X_1}$  & 191 & $\dot{C_1}$  & 149   \\
$\dot{Q_2}$  & \centering 80    & $\dot{S_2}$    & \centering 60  & $\dot{X_2}$  & 101 & $\dot{C_2}$  & 140   \\
            &       & $\dot{I}$     & 197 & $\dot{X_3}$  & 127 &             &       \\
\midrule 
\multicolumn{8}{c}{\textbf{Individual-level Model ($T1DSim_{NN,k}^{I}$)}}           \\ \addlinespace[0.2em]  
\midrule 
\multicolumn{8}{c}{$\dot{Q}_1^I$ {[}128,64,32{]} }                                    \\ \addlinespace[0.2em] 
\bottomrule                                    
\end{tabular}

\label{table:best_archs}

\end{table}

As expected, the population-level model $T1DSim_{NN}^{P}$ closely matches the $T1DSim_{ODE}^{P}$ as indicated by error metrics calculated on the hold-out simulated dataset $\mathcal{D}^{P}_{test}$:  TIR 74.1 ± 34.1\% vs. 72.9 ± 35.2\% (P-value=\textless0.001), TAR 22.3 ± 35.3\% vs. 23.1 ± 36.3\% (P-value=\textless0.001), TBR 3.6 ± 6.8\% vs. 4.0 ± 8.2\% (P-value=\textless0.001), LBGI 1.4 ± 1.5 vs. 1.5 ± 1.7 (P-value=\textless0.001), HBGI 4.1 ± 6.0 vs. 4.2 ± 6.2 (P-value=\textless0.001) and MG 132.8 ± 46.8 mg/dL vs. 132.6 ± 48.2 mg/dL (P-value=\textless0.001). The results of verifying the dynamics learned by each sub-network $\mathbb{N}_{fi}$ in $T1DSim_{NN}^{P}$ are summarized in Table \ref{table:conformance_summary}. This table shows the maximum critical error found when testing different inputs and properties within each sub-network as well as the input that produced such error. The sub-networks corresponding to states $X_2$ and $C_1$ are fully conformant. The other sub-networks are partially conformant with insignificant critical errors. For example, the maximum critical error for  $\dot{Q_1}$ is only $2e^{-2}$ mg/dL, which is a very small error in practice. A detailed version of the conformance analysis is presented in Appendix \ref{apx:conformance}.

\begin{table}[ht]
\centering
    \caption{Results of the neural network state-space model conformance verification: Summary of the maximum critical error $\Delta \mathbf{x}_{critical}^{i}$ for each sub-network of the population-level model ($T1DSim_{AI}^{P}$). If the result in the sub-network is a checkmark (\checkmark), it indicates that the sub-network is fully conformant for all tested inputs.}
\begin{tabular}{p{12mm}p{12mm} p{12mm}p{12mm} p{12mm}p{12mm} p{9mm} p{9mm}}
\toprule
\multicolumn{8}{c}{$\Delta \mathbf{x}_{critical}^{i}$}  \\ \addlinespace[0.2em]  
\midrule 
\multicolumn{2}{c}{\parbox[c]{20mm}{\centering\textbf{Glucose\\kinetics}}} & 
\multicolumn{2}{c}{\parbox[c]{20mm}{\centering\textbf{Insulin\\kinetics}}} & 
\multicolumn{2}{c}{\parbox[c]{20mm}{\centering\textbf{Insulin\\dynamics}}} & 
\multicolumn{2}{c} {\parbox[c]{20mm}{\centering\textbf{Carbohydrate\\absorption}}} \\ \addlinespace[0.5em]  
\parbox[c]{12mm}{\centering Q1\\ \scriptsize[mg/dL]}  & \parbox[c]{12mm}{\centering $2e^{-2}$ }   & \parbox[c]{12mm}{\centering S1\\ \scriptsize[mU/kg]}  & \parbox[c]{12mm}{\centering $1e^{-2}$ }  & \parbox[c]{12mm}{\centering X1\\ \scriptsize[min$^{-1}$]} & \parbox[c]{12mm}{\centering $5e^{-9}$ } & \parbox[c]{12mm}{\centering C1\\ \scriptsize[-]} & \parbox[c]{12mm}{\centering \checkmark }  \\ \addlinespace[1em]  
\parbox[c]{12mm}{\centering Q2\\ \scriptsize[mmol/kg]} & \parbox[c]{12mm}{\centering $2e^{-5}$ }   & \parbox[c]{12mm}{\centering S2\\ \scriptsize[mU/kg]}   & \parbox[c]{12mm}{\centering $9e^{-6}$ } & \parbox[c]{12mm}{\centering X2\\ \scriptsize[min$^{-1}$]} & \parbox[c]{12mm}{\centering \checkmark } & \parbox[c]{12mm}{\centering C2\\ \scriptsize[$\frac{mmol}{kg\cdot min}$]} & \parbox[b]{12mm}{\centering $2e^{-8}$ }  \\ \addlinespace[1em]
            &       & \parbox[c]{12mm}{\centering I\\ \scriptsize[mU/L]} & \parbox[c]{12mm}{\centering $2e^{-5}$ } & \parbox[c]{12mm}{\centering X3\\ \scriptsize[unitless]} & \parbox[c]{12mm}{\centering $4e^{-7}$ } &             &       \\ \addlinespace[0.4em]
\bottomrule                                    
\end{tabular}

\label{table:conformance_summary}

\end{table}

\renewcommand{\arraystretch}{1.2}
\setlength{\tabcolsep}{2pt}

\begin{table}[ht!]
\centering
\caption{Glucose outcomes for 5-hour sequences across 394 digital twins, comparing actual data, the NN-based population-level model, the ODE-based population-level model, the NN-based digital twins, the ODE-based digital twins, and the BayesianOpt-based digital twins.
Values are presented as mean$\pm$SD (p-value). Equivalence margins used for the test: $\delta_{TIR} = 5\%$, $\delta_{TAR} = 5\%$, $\delta_{TBR} = 1\%$, $\delta_{HBGI} = 1$, $\delta_{LBGI} = 0.5$ and  $\delta_{MG} = 10$ mg/dL. Equivalence is concluded only if the TOST p-value is $<$ 0.05 (i.e., both one-sided tests are significant).}

\begin{tabularx}{\textwidth}{
>{\centering\arraybackslash}p{10mm} 
>{\centering\arraybackslash}p{24mm} 
>{\centering\arraybackslash}p{24mm} 
>{\centering\arraybackslash}p{24mm} 
>{\centering\arraybackslash}p{24mm} 
>{\centering\arraybackslash}p{24mm} 
>{\centering\arraybackslash}p{24mm}}

\toprule
 & \multicolumn{1}{c}{\parbox[c][2.7em]{10mm}{\centering Actual}} 
 & \multicolumn{1}{c}{\parbox[c][2.7em]{22mm}{\centering ODE-based\\ Population-level\\(Control)}} 
 & \multicolumn{1}{c}{\parbox[c][2.7em]{22mm}{\centering ODE-based\\ Digital Twins\\(Control)}} 
 & \multicolumn{1}{c}{\parbox[c][5.1em]{22mm}{\centering Bayesian Opt-based\\ Digital Twins\\(Control)}} 
 & \multicolumn{1}{c}{\parbox[c][2.7em]{22mm}{\centering NN-based\\ Population-level\\(This work)}} 
 & \multicolumn{1}{c}{\parbox[c][2.7em]{22mm}{\centering NN-based\\ Digital Twins\\(This work)}} \\
\midrule

\multicolumn{1}{c}{\parbox[c][2.7em]{10mm}{\centering \textbf{TIR} [\%]}} & 74.4 $\pm$ 15.4 & 66.4 $\pm$ 11.5 \newline \small(1.000) & 64.7 $\pm$ 19.6 \newline \small(1.000) & 71.9 $\pm$ 12.9 \newline \textbf{\small(\textless0.001)} & 65.6 $\pm$ 13.9 \newline \small(1.000) & 75.1 $\pm$ 21.2 \textbf{\small(\textless0.001)} \\

\multicolumn{1}{c}{\parbox[c][2.7em]{10mm}{\centering \textbf{TAR} [\%]}} & 22.6 $\pm$ 15.9 & 27.7 $\pm$ 12.4 \newline \small(0.56) & 29.3 $\pm$ 22.6 \newline \small(0.999) & 21.0 $\pm$ 14.6 \newline \textbf{\small(\textless0.001)} & 33.1 $\pm$ 14.0 \newline \small(1.000) & 22.4 $\pm$ 22.0 \newline \textbf{\small(\textless0.001)} \\

\multicolumn{1}{c}{\parbox[c][2.7em]{10mm}{\centering \textbf{TBR} [\%]}} & 3.0 $\pm$ 3.3 & 5.9 $\pm$ 6.2 \newline \small(1.000) & 6.0 $\pm$ 8.3 \newline \small(1.000) & 7.1 $\pm$ 6.3  \newline \small(1.000) & 1.3 $\pm$ 1.9 \newline \small(1.000) & 2.5 $\pm$ 5.2 \newline \textbf{\small(0.022)} \\ 

\multicolumn{1}{c}{\parbox[c][2.7em]{10mm}{\centering \textbf{LBGI}}} & 0.9 $\pm$ 0.7 & 1.7 $\pm$ 1.8 \newline \small(1.000) & 1.6 $\pm$ 2.1 \newline \small(0.992) & 1.9 $\pm$ 1.7 \newline \small(1.000) & 0.4 $\pm$ 0.4 \newline \small(0.251) & 0.7 $\pm$ 1.0 \textbf{\small(\textless0.001)} \\

\multicolumn{1}{c}{\parbox[c][2.7em]{10mm}{\centering \textbf{HBGI}}} & 5.3 $\pm$ 4.1 & 6.5 $\pm$ 2.9 \newline \small(0.944) & 6.8 $\pm$ 5.8 \newline \small(1.000) & 4.9 $\pm$ 3.2  \newline \textbf{\small(\textless0.001)} & 7.6 $\pm$ 3.2 \newline \small(1.000) & 5.3 $\pm$ 5.6 \textbf{\small(\textless0.001)} \\

\multicolumn{1}{c}{\parbox[c][2.7em]{10mm}{\centering \textbf{MG} \newline [mg/dL]}} & 147.2 $\pm$ 25.5 & 153.2 $\pm$ 20.6 \newline \textbf{\small(\textless0.001)} & 153.8 $\pm$ 39.0 \newline \textbf{\small(\textless0.001)} & 140.8 $\pm$ 24.9  \newline \textbf{\small(\textless0.001)} & 165.0 $\pm$ 19.1 \newline \small(1.000) & 149.2 $\pm$ 35.1 \textbf{\small(\textless0.001)} \\
\bottomrule
\end{tabularx}

\label{table:overall_glucoseOutcomes}

\end{table}

We created a total of 394 physiologically-constrained NN digital twins. Table \ref{table:overall_glucoseOutcomes} presents the glucose outcomes for 5-hour simulations of both NN-based and ODE-based digital twins, along with the results of the population-level models.  The NN-based digital twins consistently produce more accurate simulation results across all considered metrics when compared with ODE-based digital twins using different twinning techniques. Our results demonstrate that glucose outcomes calculated from NN-based digital twins closely match real-world glucose outcomes within clinically significant glucose ranges. The difference between simulated and observed outcomes are 0.7\%, -0.2\%, -0.5\%, -0.2, 0.0, 2.0 mg/dL for TIR, TAR, TBR, LBGI, HBGI, and MG, respectively (TIR 75.1±21.2\% vs. 74.4±15.4\% (P-value=\textless0.001), TAR 22.4±22.0\% vs. 22.6±15.9\% (P-value=\textless0.001), TBR 2.5±5.2\% vs. 3.0±3.3\% (P-value=0.022), LBGI 0.9±0.7 vs. 0.7±1.0 (P-value=\textless0.001), HBGI 5.3±4.1 vs. 5.3±5.6 (P-value=\textless0.001) and MG 145.2±25.5 mg/dL vs. 149.2±35.1 mg/dL (P-value=\textless0.001)).

\renewcommand{\arraystretch}{1.2}
\setlength{\tabcolsep}{2pt}


\begin{table}[!ht]
\centering
\caption{Performance comparison of digital twins trained with varying input configurations across three scenarios: physical activity, meals, and sleep. 
Values are presented as mean$\pm$SD (p-value). Equivalence margins used for the test: $\delta_{TIR} = 5\%$, $\delta_{TAR} = 5\%$, $\delta_{TBR} = 1\%$, $\delta_{HBGI} = 1$, $\delta_{LBGI} = 0.5$ and  $\delta_{MG} = 10$ mg/dL. Equivalence is concluded only if the TOST p-value is $<$ 0.05 (i.e., both one-sided tests are significant).}

\begin{tabularx}{\textwidth}{
>{\centering\arraybackslash}p{10mm} 
>{\centering\arraybackslash}p{21mm} 
>{\centering\arraybackslash}p{21mm} 
>{\centering\arraybackslash}p{21mm} 
>{\centering\arraybackslash}p{21mm} 
>{\centering\arraybackslash}p{21mm}
>{\centering\arraybackslash}p{21mm}
>{\centering\arraybackslash}p{21mm}}
\toprule
 & \parbox[c][2.7em]{21mm}{\centering Actual} 
 & \parbox[c][3.2em]{21mm}{\centering All\\ Individual\\ Inputs} 
 & \parbox[c][2.7em]{21mm}{\centering Heart\\ Rate} 
 & \parbox[c][2.7em]{21mm}{\centering Sleep\\ Efficiency} 
 & \parbox[c][2.7em]{21mm}{\centering Weekend\\ day} 
 & \parbox[c][2.7em]{21mm}{\centering Time-based\\ Features} 
 & \parbox[c][2.7em]{21mm}{\centering No Additional\\ Inputs} \\
\midrule

\multicolumn{8}{c}{\textbf{\textit{Scenario: Physical Activity}}} \\

\textbf{TIR} [\%] & $73.9 \pm 16.8$ & $73.1 \pm 24.4$\newline \textbf{\small(\textless 0.001)} & $73.7 \pm 25.7$\newline \textbf{\small(\textless 0.001)} & $74.1 \pm 26.3$\newline \textbf{\small(\textless 0.001)} & $73.5 \pm 26.2$\newline \textbf{\small(\textless 0.001)} & $73.3 \pm 25.4$\newline \textbf{\small(\textless 0.001)} & $73.5 \pm 26.5$ \newline \textbf{\small(\textless 0.001)} \\

\textbf{TAR} [\%] & $22.2 \pm 17.5$ & $24.2 \pm 25.4$\newline \textbf{\small(\textless 0.001)} & $24.7 \pm 26.4$\newline \textbf{\small(0.002)} & $24.8 \pm 26.8$\newline \textbf{\small(0.003)} & $25.5 \pm 26.5$\newline \textbf{\small(0.020)} & $24.6 \pm 25.9$\newline \textbf{\small(\textless 0.001)} & $25.7 \pm 26.8$ \newline \textbf{\small(0.039)} \\

\textbf{TBR} [\%] & $3.8 \pm 4.7$ & $2.7 \pm 6.5$\newline \small(0.679) & $1.6 \pm 4.2$\newline \small(1.000) & $1.0 \pm 3.6$\newline 
\small(1.000) & $1.0 \pm 3.3$\newline 
\small(1.000) & $2.1 \pm 6.8$\newline \small(0.989) & $0.8 \pm 2.9$\newline \small(\textless 1.000) \\ 

\textbf{LBGI} & 1.0 $\pm$ 1.0  & 0.7 $\pm$ 1.4 \newline \textbf{\small(0.004)} & 0.5 $\pm$ 1.0 \newline \small(0.656) & 0.4 $\pm$ 0.9 \newline \small(0.999) & 0.4 $\pm$ 0.9 \small(0.999) & 0.6 $\pm$ 1.3 \newline \small(0.383) & 0.3 $\pm$ 0.8 \small(1.000)  \\

\textbf{HBGI} & 5.1 $\pm$ 4.3 & 5.7 $\pm$ 6.1 \newline \small\textbf{(0.013)} & 5.6 $\pm$ 5.8  \newline \small\textbf{(0.002)} & 5.7 $\pm$ 6.1 \newline \small(\textbf{0.017}) & 5.8 $\pm$ 6.1 \small(0.059) & 5.7 $\pm$ 5.8 \newline \small\textbf{(0.009)} & 5.8 $\pm$ 5.9 \small(0.071) \\

\textbf{MG} \newline [mg/dL] & 145.5 $\pm$ 27.8 & 152.0 $\pm$ 38.9 \newline \textbf{\small(0.002)} & 152.7 $\pm$ 37.5 \newline \textbf{\small(0.008)} & 153.9 $\pm$ 38.5  \newline \small(0.091) &  154.8 $\pm$ 38.0 \small(0.260) & 153.1 $\pm$ 37.5 \newline \textbf{\small(0.015)} & 155.4 $\pm$ 37.3 \small(0.463) \\
\hline

\multicolumn{8}{c}{\textbf{\textit{Scenario: Meals}}} \\

\textbf{TIR} [\%] & $70.5 \pm 16.2$ & $70.3 \pm 24.8$ \newline \textbf{\small(\textless 0.001)} & $71.6 \pm 26.7$ \newline \textbf{\small(\textless 0.001)} & $72.1 \pm 26.8$ \newline \textbf{\small(\textless 0.001)} & $71.4 \pm 26.7$ \newline \textbf{\small(\textless 0.001)} & $70.7 \pm 25.7$ \newline \textbf{\small(\textless 0.001)} & $71.7 \pm 27.1$ \newline \textbf{\small(\textless 0.001)} \\

\textbf{TAR} [\%] & $26.1 \pm 17.1$ & $27.9 \pm 25.6$ \newline \textbf{\small(\textless 0.001)} & $27.2 \pm 27.1$ \newline \textbf{\small(\textless 0.001)} & $26.8 \pm 27.2$ \newline \textbf{\small(\textless 0.001)} & $27.3 \pm 27.1$ \newline \textbf{\small(\textless 0.001) }& $27.6 \pm 26.4$ \newline \textbf{\small(\textless 0.001) }& $27.3 \pm 27.4$ \newline \textbf{\small(\textless 0.001)} \\

\textbf{TBR} [\%] & $3.4 \pm 3.5$ & $1.9 \pm 4.6$ \newline \small(0.999) & $1.2 \pm 3.1$ \newline \small(1.000) & $1.1 \pm 3.2$ \newline \small(1.000) & $1.3 \pm 3.7$ \newline \small(1.000) & $1.7 \pm 4.7$ \newline \small(0.999) & $1.0 \pm 3.0$ \newline \small(1.000) \\ 

\textbf{LBGI} & $0.9 \pm 0.8$ & $0.5 \pm 1.0$ \newline \textbf{\small(0.012)} & $0.4 \pm 0.8$ \newline \small(0.891) & $0.4 \pm 0.9$ \newline \small(0.953) & $0.4 \pm 1.0$ \newline \small(0.779) & $0.5 \pm 1.0$ \newline \small(0.114) & $0.3 \pm 0.8$ \newline \small(0.987) \\

\textbf{HBGI} & $5.9 \pm 4.3$ & $6.6 \pm 6.6$ \newline \textbf{\small(0.028)} & $6.4 \pm 6.4$ \newline \textbf{\small(\textless 0.001)} & $6.4 \pm 6.7$ \newline \textbf{\small(\textless 0.001)} & $6.4 \pm 6.7$ \newline \textbf{\small(0.001) }& $6.4 \pm 6.4$ \newline \textbf{\small(\textless 0.001) }& $6.4 \pm 6.5$ \newline \textbf{\small(\textless 0.001)} \\

\textbf{MG} \newline [mg/dL] & $151.2 \pm 26.9$ & $158.6 \pm 39.7$ \newline \textbf{\small(0.005)} & $158.3 \pm 38.8$ \newline \textbf{\small(0.002)} & $158.1 \pm 40.5$ \newline \textbf{\small(0.002)} & $158.4 \pm 40.2$ \newline \textbf{\small(0.003) }& $157.9 \pm 39.0$ \newline \textbf{\small(\textless 0.001) }& $158.8 \pm 39.3$ \newline \textbf{\small(0.009)} \\
\hline

\multicolumn{8}{c}{\textbf{\textit{Scenario: Sleep}}} \\

\textbf{TIR} [\%] & $76.3 \pm 18.8$ & $77.9 \pm 22.0$ \newline \textbf{\small(\textless 0.001) }& $80.0 \pm 22.5$ \newline \small(0.064) & $79.4 \pm 23.0$ \newline \textbf{\small(0.011)} & $81.7 \pm 21.8$ \newline \small(0.695) & $78.7 \pm 23.0$ \newline \textbf{\small(\textless 0.001)} & $81.3 \pm 22.7$ \newline \small(0.489) \\

\textbf{TAR} [\%] & $21.1 \pm 18.6$ & $18.4 \pm 22.0$ \newline \textbf{\small(0.002)} & $17.2 \pm 22.5$ \newline \small(0.069) & $17.8 \pm 23.1$ \newline \textbf{\small(0.014)} & $15.6 \pm 21.7$ \newline \small(0.720) & $18.7 \pm 23.2$ \newline \textbf{\small(\textless 0.001)} & $16.2 \pm 22.8$ \newline \small(0.430) \\

\textbf{TBR} [\%] & $2.6 \pm 4.3$ & $3.7 \pm 8.3$ \newline  \textbf{\small(0.001)} & $2.8 \pm 7.0$ \newline  \textbf{ \small(\textless 0.001)}& $2.8 \pm 7.2$ \newline  \textbf{\small(0.017)} & $2.6 \pm 7.0$ \newline  \textbf{\small(0.004)} & $2.5 \pm 6.7$ \newline  \textbf{\small(0.005)} & $2.5 \pm 6.9$ \newline \textbf{\small(0.009)} \\

\textbf{LBGI} & $0.8 \pm 1.0$ & $1.1 \pm 1.5$ \newline \textbf{\small(0.001)} & $1.0 \pm 1.4$ \newline \textbf{\small(0.001)} & $0.9 \pm 1.3$ \newline \textbf{\small(\textless0.001)} & $0.9 \pm 1.3$ \newline \textbf{\small(\textless0.001)} & $0.9 \pm 1.2$ \newline \textbf{\small(\textless0.001)} & $0.9 \pm 1.3$ \newline \textbf{\small(\textless0.001)} \\

\textbf{HBGI} & $5.1 \pm 4.6$ & $4.3 \pm 5.2$ \newline\small(0.132) & $3.8 \pm 4.7$ \newline \small(0.943) & $4.1 \pm 5.1$ \newline \small(0.568) & $3.7 \pm 5.1$ \newline \small(0.992)& $4.2 \pm 5.0$ \newline \small(0.179) & $3.7 \pm 4.8$ \newline \small(0.996) \\

\textbf{MG} \newline [mg/dL] & $146.0 \pm 29.2$ & $138.3 \pm 36.6$ \newline \textbf{\small(0.039)} & $135.9 \pm 35.2$ \newline \small(0.535) & $137.9 \pm 36.2$ \newline \small(0.056) & $134.8 \pm 36.3$ \newline \small(0.859) &  $138.8 \pm 36.6$ \newline \textbf{\small(0.008)} &  $135.1 \pm 35.8$ \newline \small(0.775) \\
\bottomrule

\end{tabularx}

\label{table:inputsContributions}

\end{table}

Table \ref{table:inputsContributions} summarizes the similarity between simulated and actual glucose profiles under varying input configurations across three scenarios: physical activity, meals, and sleep. Digital twins that included all available inputs produced glucose outcomes more closely aligned with real-world data. For physical activity and meal inputs, when all variables are included, all metrics passed the equivalence test except for TBR; for sleep, all metrics passed except for HBGI. In other input configurations, the behavior was not the same, as they failed at least two of the tests. For example, in the physical activity scenario, when no additional inputs were included, the MG, TBR, LBGI, and HBGI were not within the predefined equivalence margins. However, when models were trained using heart rate or time-based features, MG and HBGI became equivalent to real-world outcomes. Furthermore, when all individual-level inputs were included, LBGI also reached equivalence. This is a significant finding, as it indicates that incorporating these inputs adds value and improves the accuracy of simulated outcomes. 

\begin{figure*}[h]
\centering
\includegraphics[width=\textwidth]{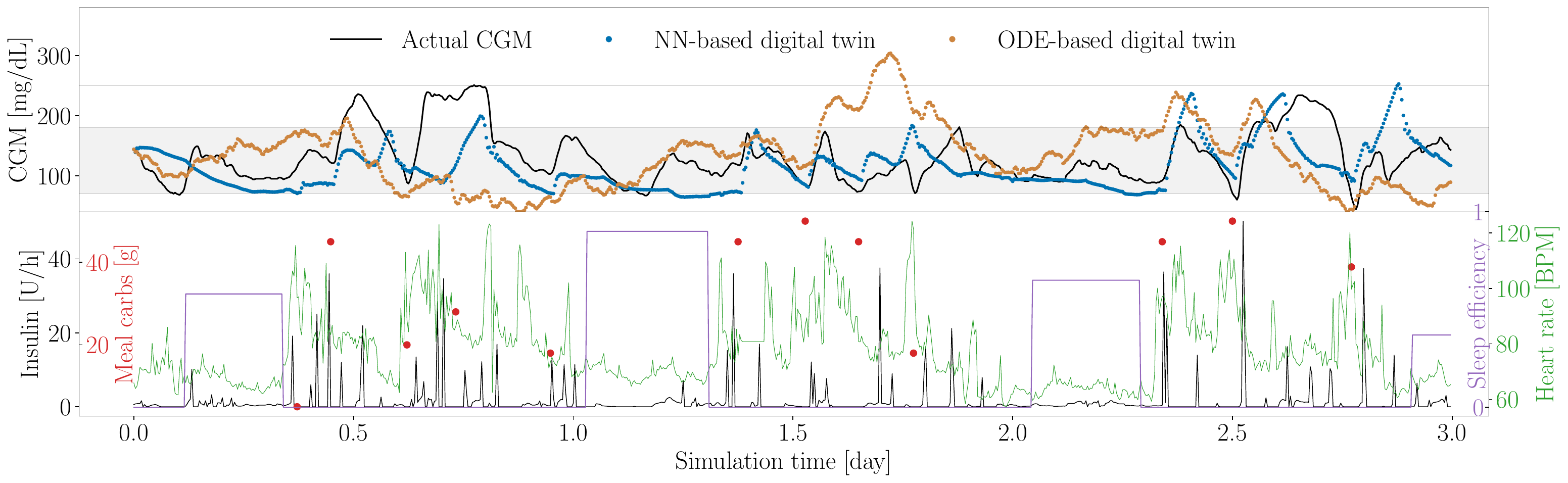}
\caption{Example of a 3-day glucose trace from the actual CGM vs. simulated (NN-based digital twin vs. ODE-based digital twin) (Top panel). Glucose management scenario used for the simulation, which includes endogenous insulin, amount of carbohydrates, and heart rate and sleep efficiency (Bottom panel). TBR, TAR, and TIR for the actual trace are 1.9\%, 15.0\%, and 83.1\%, respectively. For the NN-based digital twin, the values are 4.7\%, 6.8\%, and 88.4\%, while for the ODE-based digital twin, they are 16.6\%, 20.8\%, and 62.6\%.}
\label{fig:pilot_error}
\end{figure*}

\begin{figure*}[!h]
\centering
\includegraphics[width=\textwidth]{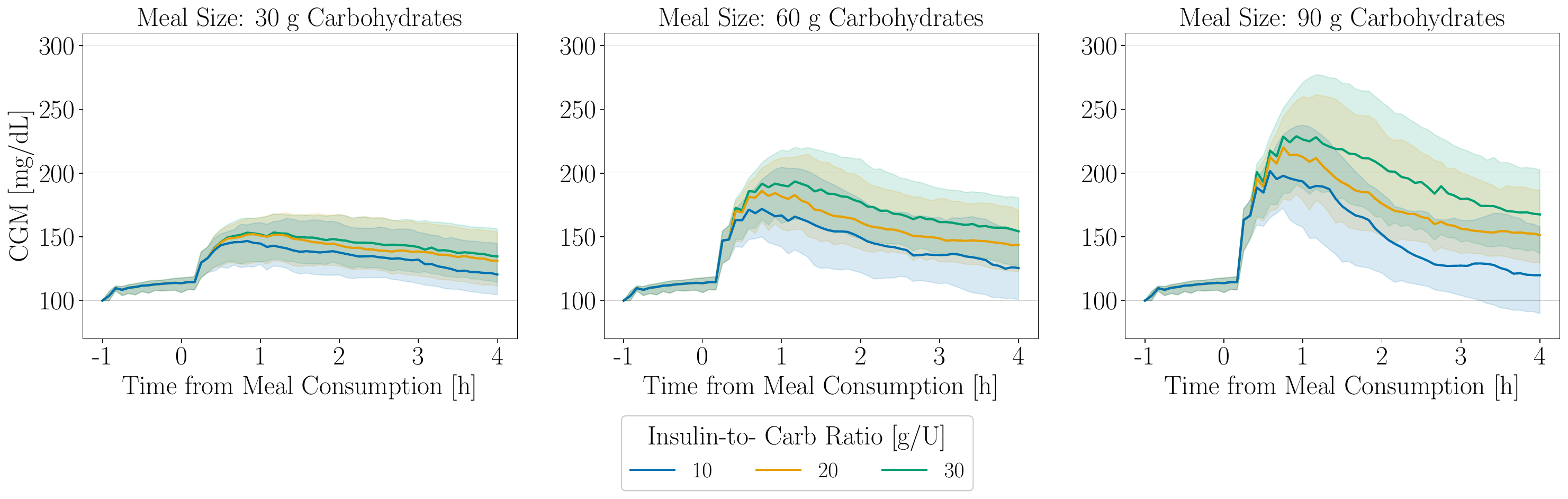}
\caption{Simulated responses of physiologically-constrained NN digital twins to three meal scenarios (from left to right: 30 g, 60 g, and 90 g of carbohydrate consumption). For each meal scenario, insulin boluses were simulated at three insulin-to-carbohydrate ratios (10 g/U, 20 g/U, and 30 g/U). Results for each simulated scenario in the set of physiologically-constrained NN digital twins are shown as the median (thick line) and interquartile range (shaded area). Note that increased carbohydrate consumption leads to higher postprandial glucose, while higher insulin-to-carbohydrate ratios reduce postprandial glucose for each bolus tested. \vspace{1.5em}}
\label{fig:twinning_effect}
\end{figure*}

Fig. \ref{fig:pilot_error} shows an illustrative example of a 3-day simulation of an NN-based digital twin (this work) and an ODE-based digital twin (control) as well as the actual CGM values. In this example, our approach exhibits a better match in glucose outcomes compared to the ODE-based digital twin. Additionally, it consistently captures expected physiological dynamics, such as an increase in glucose following meal inputs and a reduction in glucose in response to elevated heart rate, possibly due to physical activity. Moreover, Fig. \ref{fig:twinning_effect} qualitatively illustrates the ability of the NN-based digital twins to adapt to varying inputs, showcasing their potential for individual-specific optimization of factors such as insulin-to-carbohydrate ratios. Furthermore, the results demonstrate variability in the response among the digital twins. The results are presented in Figure \ref{fig:twinning_effect}

Finally, our point-based RMSE for a 30-, 60- and 120-minute prediction horizon is 27±7 mg/dL, 41±12 mg/dL and 54±18 mg/dL, respectively, which is in alignment with other state-of-the-art models \cite{Kushner_Sankaranarayanan_Breton_2020, Prendin_Pavan_Cappon_Del_2023,Kushner_Breton_Sankaranarayanan_2020}.
These results demonstrate that our proposed digital twins can function not only as a simulation tool but also as effective short-term glucose predictors.

\section{Discussion}\label{sec:discussion}
We described a framework for constructing physiologically-constrained NN digital twins capable of replicating the glucose dynamics of people living with T1D. Our proposed approach combines a population-level model $T1DSim_{NN}^{P}$ that is verified to conform to known glucose-insulin dynamics within very low critical errors bounds with an individual-level model $T1DSim_{NN}^{I}$ that captures residual dynamics not modeled by $T1DSim_{NN}^{P}$. The digital twins constructed in this work not only demonstrated good agreement with actual glucose values when simulating meal, insulin, exercise, and sleep scenarios (Table \ref{table:overall_glucoseOutcomes}), but also demonstrated their potential as a tool for simulating multiple interventions and optimizing treatment for individuals living with T1D (Fig. \ref{fig:twinning_effect}). 

Using our proposed framework, digital twins can be constantly fine-tuned or extended by adding new inputs as new data from the twined individuals become available. This means the glucose dynamics of each digital twin can evolve along with the real-world twin using simple and fast methods involving gradient descent optimization techniques in contrast to the complex design and model identification techniques used for developing mechanistic models.

Our approach provide advances of significant utility by enabling long-term simulations across a multi-hour window, these digital twins can enable \textit{in-silico} experiments that can enhance treatment strategies for individuals with T1D. We expect that the accuracy of the model will be sufficient for use in providing decision support around alternate treatment strategies such as alternative carbohydrate ratios or other settings within diabetes devices, but this will need to be tested in future research.

A natural question arising from this work is why we chose to adopt a NN state-space model design instead of simply creating a set of NN-based digital twins derived from the $T1DSim_{ODE}$ model. Our approach provides a physiologically meaningful and flexible framework for modeling the glucoregulatory system in T1D. Unlike traditional ODE-based models, the NN state-space model offers several advantages. First, it allows for continuous computationally efficient training as new data becomes available, ensuring adaptability to evolving clinical datasets. Second, it facilitates seamless integration into decision support systems, enhancing its applicability in personalized diabetes management. Third, the framework can be easily extended by incorporating additional compartments to account for other physiological factors affecting glucose regulation. These advantages position the NN state-space approach as a robust alternative for the creation of digital twins that can continuously evolve alongside their real-world counterparts.

While the number of parameters to be adjusted is larger than other approaches, we mitigate the risk of overfitting by the availability of extensive training data. Specifically, the population-level model contains 6,782 trainable parameters, representing less than 0.7\% of the approximately 1 million 5-hour sequences used for training. For the individual-level models, although they are more complex relative to the limited training data available, have significant reason to believe the observed improvements are not solely due to the model's complexity. Rather, factors such as the architecture design, the quality of the data, and the use of regularization techniques like weight decay play a critical role in preventing overfitting. Moreover, a key advantage of these digital twins is their ability to model additional features not included in the deterministic ODE model while continuing to train as more individual data becomes available, enhancing their performance and adaptability over time.

It is noteworthy that our initial approach to creating the individual-level models involved modeling the residual errors between the actual glucose data and the population-level model. However, this approach proved ineffective, as it tended to learn an average mean error. This was likely due to factors such as the intrinsic dependency on the duration of the population-level simulation, which made it challenging to capture the unmodeled phenomena driven by factors like physical activity or insulin sensitivity. We found our current approach improves robustness, as it directly adjusts each digital twin's simulation to reflect changes in glucose dynamics, enabling a more accurate representation of individual variability.

\section{Conclusion}\label{sec:conclusion}
This work presents a framework for constructing physiologically-constrained NN digital twins that can simulate glucose dynamics of individuals with T1D. We developed a novel NN state-space model architecture that allows for observability and interpretability of simulation outputs. This model adheres to known glucose-insulin dynamics as verified by our conformance verification analysis. When augmented with individual-level data, the resulting adaptive digital twin models can capture both inter- and intra-individual variability and incorporate various factors influencing glucose response, such as sleep and physical activity. We show that incorporating this data significantly improves simulation accuracy compared to digital twins trained with limited or varying input configurations, and that our framework outperforms ODE-based digital twins, highlighting the benefits of data-driven modeling combined with comprehensive input integration. Unlike traditional mechanistic models constrained by fixed parameters, our approach offers the flexibility to continuously adapt to new data, tasks, or architectures, enabling a dynamic and scalable framework for personalized modeling.

\section*{Acknowledgments}
This research was funded by Breakthrough T1D, formerly JDRF (grant 2-SRA-2022-1273-S-B).

\clearpage
\begin{appendices}

\section{Comparison of the proposed framework with existing digital twin approaches}\label{apx:related_work}

\begin{table}[h]
\centering
    \caption{Comparison of the proposed framework with existing digital twin approaches, highlighting architecture, learning process, inputs, code availability and data used for parameter identifications.}
\begin{tabular}{c c c}
\toprule
\multicolumn{3}{c}{\textbf{Digital Twins Framework for T1D}}  \\ \addlinespace[0.2em]  
\midrule 
 & \textbf{Our Approach} & 
\textbf{Other Approaches} \\ \addlinespace[0.7em]  
\parbox[c]{ 18mm}{\centering \textit{Architecture}}  & \parbox[c]{55mm}{First methodology using a neural network state space model architected based on an ODE-based mechanistic model} & \parbox[c]{55mm}{None are based on neural networks \cite{G_2023,Colmegna_2020,Deichmann_2023,Goodwin_2020,Haidar_2013,Hughe_2021,Young_2023,Visentin_2016}} \\ \addlinespace[1em] 
\parbox[c]{ 18mm}{\centering \textit{Model fitting}}  & \parbox[c]{55mm}{Gradient descent} & \parbox[c]{55mm}{Markov chain Monte Carlo \cite{G_2023, Haidar_2013, Young_2023}; Latin-Hypercube Sampling \cite{Colmegna_2020};Parameters obtained from literature \cite{Deichmann_2023}; Gradient-based method \cite{Goodwin_2020}; Least-square Fitting \cite{Hughe_2021};Bayesian Maximum a Posteriori\cite{Visentin_2016}} \\ \addlinespace[1em] 
\parbox[c]{ 18mm}{\centering \textit{Simulations}}  & \parbox[c]{55mm}{Digital twins can simulate real-world scenarios with no need for segmentation of events such as meals or exercise} & \parbox[c]{55mm}{Digital twins are created based on specific events such exercise \cite{Young_2023}, or constrained meal scenarios \cite{Colmegna_2020,Deichmann_2023,Haidar_2013,Visentin_2016}}. \\ \addlinespace[1em] 
\parbox[c]{ 18mm}{\centering \textit{Code \\ Availabilty}}  & \parbox[c]{55mm}{ Our framework code is available on \href{https://github.com/XXX/XXX}{GitHub}} & \parbox[c]{55mm}{Only \cite{G_2023,Deichmann_2023} have public repositories with the code} \\ \addlinespace[1em] 
\parbox[c]{ 18mm}{\centering \textit{Inputs}}  & \parbox[c]{55mm}{ It uses insulin, carbohydrate intake, heart rate, sleep data, and time-based contextual features.} & \parbox[c]{55mm}{ Usually include only insulin and carbohydrate intake \cite{G_2023,Colmegna_2020,Goodwin_2020,Haidar_2013,Hughe_2021,Visentin_2016}, only a few have included heart rate information \cite{Deichmann_2023,Young_2023}} \\ \addlinespace[1em] 
\parbox[c]{ 18mm}{\centering \textit{Flexibility}}  & \parbox[c]{55mm}{ The framework is capable of (1) incorporating new inputs that may affect glucose levels, (2) modifying the loss function to adapt to individual needs (e.g., high glucose variability), or (3) computationally efficient continuous training as more data becomes available.} & \parbox[c]{55mm}{ No other digital twin approach offers this level of flexibility. Some existing methods are constrained to be fitted to each event, making them computationally intensive and impractical for real-time applications \cite{G_2023}.} \\ \addlinespace[1em] 
\bottomrule                                    
\end{tabular}

\label{table:digitalTwinReview}

\end{table}

\clearpage

\section{Description of the datasets used for models' development and testing}\label{apx:dataset}

\begin{table}[h]
\centering
\caption{Description of the datasets used for models' development and testing }
    \begin{tabular}{m{45mm}p{30mm}m{30mm}} 
    \toprule
    \multirow{2}{*}{\textbf{Characteristic}} & \multicolumn{2}{c}{\textbf{Dataset}} \\ \cmidrule(lr){2-3} 
    \multicolumn{1}{c}{}                     & \multicolumn{1}{c}{\textbf{Simulated}}   & \multicolumn{1}{c}{\textbf{T1DEXI Main Study}} \\ 
    \midrule
    \textbf{Use}                                                 & \multicolumn{1}{c}{Population} & \multicolumn{1}{c}{Personalization} \\ 
    \textbf{Participants, N  }                                   & \multicolumn{1}{c}{}                    & \multicolumn{1}{c}{394} \\
    \multicolumn{3}{l}{\textbf{Demographics}} \\
    \multicolumn{3}{l}{\hspace{1mm}\textit{Biological sex, N}} \\
    \hspace{2mm}Female (Male)                                    & \multicolumn{1}{c}{}                    & \multicolumn{1}{c}{295 (99)} \\
    \multicolumn{3}{l}{\hspace{1mm}\textit{Age, years}} \\
    \hspace{2mm}Mean$\pm$SD                                        & \multicolumn{1}{c}{}                    & \multicolumn{1}{c}{37$\pm$14} \\
    \multicolumn{3}{l}{\hspace{1mm}\textit{BMI, kg/m}$^2$}\\
    \hspace{2mm}Mean$\pm$SD                                        & \multicolumn{1}{c}{}                    & \multicolumn{1}{c}{25$\pm$4}  \\
    \multicolumn{3}{l}{\hspace{1mm}\textit{Insulin therapy, N}} \\
    \hspace{2mm}Pump                                             & \multicolumn{1}{c}{}                    & \multicolumn{1}{c}{179} \\
    \hspace{2mm}Closed loop                                      & \multicolumn{1}{c}{}                    & \multicolumn{1}{c}{215} \\ \addlinespace[0.2em]
    \multicolumn{3}{l}{\textbf{Overall glucose control, Mean {[}Min - Max{]} at participant level}} \\ \addlinespace[0.2em]
    \multicolumn{3}{l}{\hspace{1mm} \textit{CGM between 70-180 mg/dL, }\%} \\
    \multicolumn{1}{c}{Mean}                                            & \multicolumn{1}{c}{73.0}              & \multicolumn{1}{c}{75.2}  \\
    \multicolumn{1}{c}{\hspace{2mm} [Min - Max]}                                          & \multicolumn{1}{c}{{[}70.0 - 76.6{]}}  & \multicolumn{1}{c}{{[}20.7 - 99.1{]}} \\
    \multicolumn{3}{l}{\hspace{1mm}\textit{CGM \textgreater{}180 mg/dL,} \%} \\
    \multicolumn{1}{c}{Mean}                                            & \multicolumn{1}{c}{23.2}              & \multicolumn{1}{c}{21.8} \\
    \multicolumn{1}{c}{\hspace{2mm} [Min - Max]}                                          & \multicolumn{1}{c}{{[}20.1 - 27.0{]}}  & \multicolumn{1}{c}{{[}0.1 - 79.3{]}} \\
    \multicolumn{3}{l}{\hspace{1mm}\textit{CGM \textless{}70 mg/dL,} \%} \\ 
    \multicolumn{1}{c}{Mean}                                            & \multicolumn{1}{c}{3.7}               & \multicolumn{1}{c}{3.0} \\
    \multicolumn{1}{c}{\hspace{2mm} [Min - Max]}                                          & \multicolumn{1}{c}{{[}2.8 - 6.1{]}}    & \multicolumn{1}{c}{{[}0.0 - 17.0{]}} \\
    \bottomrule
    \end{tabular}
\label{table:descriptionDatasets}
\end{table}

\section{Robust scaler} \label{sec:robustScaler}

Robust scaling involves subtracting the median of the distribution of all values that a given compartment can take in the training dataset from the unscaled state value, and then dividing by the distribution interquartile range. For example, to scale the  $Q_1$ compartment state value at time $t$, we used Equation \ref{eq:robust_scaler}. Scaling constants for all system's states were calculated from the training dataset and stored for subsequent use during the model validation and testing phases.

In Equation \ref{eq:robust_scaler}, ${Q_1}{(t)_{sc}}$ is the scaled version of ${{Q_1}(t)}$; ${{Q_{1,(50)}}}$ is the $50^{th}$ percentile or median value of the distribution of all values that the $Q_1$ compartment can take in the training set; ${{Q_{1,(25)}}}$ and ${{Q_{1,(75)}}}$ are the ${{25}^{th}}$ and ${{75}^{th}}$ percentiles, respectively. 

\begin{equation}
    {Q_1}{(t)_{sc}} = \frac{{{Q_1}(t) - {Q_{1,(50)}}}}{{{Q_{1,(75)}} - {Q_{1,(25)}}}}
\label{eq:robust_scaler}
\end{equation}

\section{Conformance verification results}\label{apx:conformance}

\begin{table}[ht]
\centering
    \caption{Conformance verification results of each sub-network of the population-level model ($T1DSim_{AI}^{P}$)}
    \label{table:conformance}
\begin{tabular}{ccccc}
\toprule
\multirow{2}{*}{\parbox[c]{13mm}{\centering \textbf{Neural \\Network}}}      & \multirow{2}{*}{\parbox[c]{10mm}{\centering \textbf{Input Tested}}}      & \multirow{2}{*}{\parbox[c]{10mm}{\centering \textbf{Property Tested}}} & \multicolumn{2}{c}{\parbox[c]{40mm}{\centering $\Delta \mathbf{x}_{critical}^{i}$}} \\ \addlinespace[0.5em] \cmidrule(lr){4-5} 
& & & \multicolumn{1}{c}{\parbox[c][2.3em]{25mm}{\centering \textbf{Training region}}} & \multicolumn{1}{c}{\parbox[c][2.3em]{25mm}{\centering \textbf{Generalization region}}} \\ \addlinespace[0.3em]  \addlinespace[0.3em] 
\hline
\multirow{5}{*}{\parbox[c]{25mm}{\centering \textbf{Q1}\\\textbf{[mg/dL]}}} & 
                    \multirow{2}{*}{X1}     & $\downarrow$         & \checkmark             & $1e^{-12}$   \\
                    &                       & $\uparrow$           & $\mathbf{2e^{-2}}$       & $5e^{-3}$    \\ \addlinespace[0.3em] 
                    & X3                    & $\downarrow$         & $2e^{-4}$       & \checkmark           \\ \addlinespace[0.3em] 
                    & \multirow{2}{*}{Q1}   & $\downarrow$         & $8e^{-6}$       & $2e^{-5}$     \\
                    &                       & $\uparrow$           & $4e^{-5}$       & \checkmark    \\ \addlinespace[0.3em] 
                    & Q2                    & $\uparrow$           & $6e^{-13}$      & \checkmark           \\ \addlinespace[0.3em] 
                    & C2                    & $\uparrow$           & $1e^{-13}$      & \checkmark           \\ 
\hline 
\multirow{8}{*}{\parbox[c]{25mm}{\centering \textbf{Q2}\\ \textbf{[mmol/kg]}}} & 
                    \multirow{2}{*}{X1}     & $\uparrow$           & $6e^{-6}$       & \checkmark          \\
                    &                       & $\downarrow$         & $9e^{-6}$       & $1e^{-5}$    \\ \addlinespace[0.3em] 
                    & \multirow{2}{*}{X2}   & $\downarrow$         & $1e^{-5}$       & $\mathbf{2e^{-5}}$    \\
                    &                       & $\uparrow$           & $\mathbf{2e^{-5}}$       & $2e^{-6}$    \\ \addlinespace[0.3em] 
                    & \multirow{2}{*}{Q1}   & $\uparrow$           & \checkmark      & \checkmark    \\
                    &                       & $\downarrow$         & \checkmark      & $2e^{-7}$     \\ \addlinespace[0.3em] 
                    & \multirow{2}{*}{Q2}   & $\downarrow$         & \checkmark             & \checkmark           \\
                    &                       & $\uparrow$           & $3e^{-7}$       & $4e^{-7}$     \\
\hline 
\multirow{2}{*}{\parbox[c]{25mm}{\centering \textbf{S1}\\ \textbf{ [mU/kg]}}} & S1                 & $\downarrow$         & $1e^{-5}$            & \checkmark        \\ \addlinespace[0.3em] 
                    & $u_I$     & $\uparrow$           & $\mathbf{1e^{-2}}$   & \checkmark    \\
\hline 
\multirow{2}{*}{\parbox[c]{25mm}{\centering \textbf{S2}\\ \textbf{ [mU/kg]}}} & S1                 & $\uparrow$           & \checkmark  & $\mathbf{9e^{-6}}$     \\ \addlinespace[0.3em] 
                    & S2                 & $\downarrow$         & \checkmark                   & \checkmark                   \\
\hline 
\multirow{2}{*}{\parbox[c]{25mm}{\centering \textbf{I}\\ \textbf{ [mU/L]}}}  & S2                 & $\uparrow$           & \checkmark     & \checkmark          \\ \addlinespace[0.3em] 
                    & I                  & $\downarrow$         & \checkmark            & $\mathbf{2e^{-5}}$          \\
\hline 
\multirow{2}{*}{\parbox[c]{25mm}{\centering \textbf{X1}\\ \textbf{ [min$^-1$]}}} & I                  & $\uparrow$           & $7e^{-19}$        & \checkmark                \\ \addlinespace[0.3em] 
                    & X1                 & $\downarrow$         & \checkmark        & $\mathbf{5e^{-9}}$    \\
\hline 
\multirow{2}{*}{\parbox[c]{25mm}{\centering \textbf{X2}\\ \textbf{ [min$^-1$]}}} & I                  & $\uparrow$          & \checkmark    & \checkmark    \\ \addlinespace[0.3em] 
                    & X2                 & $\downarrow$         & \checkmark             & \checkmark                          \\
\hline 
\multirow{2}{*}{\parbox[c]{25mm}{\centering \textbf{X3}\\ \textbf{ [unitless]}}} & I                  & $\uparrow$           & \checkmark        & \checkmark             \\ \addlinespace[0.3em] 
                    & X3                 & $\downarrow$         & $1e^{-7}$               & $\mathbf{4e^{-7}}$      \\
\hline 
\multirow{2}{*}{\parbox[c]{25mm}{\centering \textbf{C2 [mmol/kg/min]}}} & C1                 & $\downarrow$         & \checkmark          & $7e^{-12}$    \\ \addlinespace[0.3em] 
                    & C2                 & $\uparrow$           & $9e^{-9}$     & $\mathbf{2e^{-8}}$     \\
\hline 
\multirow{2}{*}{\parbox[c]{25mm}{\centering \textbf{C1}\\ \textbf{ [-]}}} & C1                 & $\downarrow$         & \checkmark     & \checkmark                  \\ \addlinespace[0.3em] 
                    & $u_{carbs}$ & $\uparrow$           &  \checkmark                     & \checkmark         \\
\bottomrule
\end{tabular}
\end{table}

For each sub-network $\mathbb{N}_{fi}$, we performed the conformance analyses over two regions:
\begin{enumerate}
    \item \textbf{Training region:} This region considers minimum and maximum test inputs values within the distribution of values used for training.
    \item \textbf{Generalization region:} This region considers values outside the training distribution to demonstrate model generalization.
\end{enumerate}
For both experiments, we set $\delta$ as 10 times the minimum change possible of the input to be tested. It is noteworthy that for some inputs tested, we had to test two properties. The reason is due to scaling: because we rescaled the values of the states using a robust scaler, we had both negative and positive values, which affected the properties for ODEs where some components are defined as the interactions between two states (e.g., the interaction $-X_1(t)Q_1(t)$). Table \ref{table:conformance} shows the detailed results of the conformance verification for each sub-network of the population-level model ($T1DSim_{NN}^{P}$).

\clearpage

\end{appendices}

\bibliographystyle{unsrt}
\bibliography{references}  

\end{document}